\documentclass[11pt,a4paper]{article}

\usepackage[utf8]{inputenc}
\usepackage[T1]{fontenc}
\usepackage[english]{babel}
\usepackage{multirow}
\usepackage{subcaption}
\usepackage{tikz}
\usepackage{amsfonts, amsmath, amssymb}
\usepackage{newpxtext,newpxmath}
\usepackage{algorithm}
\usepackage{algorithmic}
\usepackage{hyperref}
\usepackage{cleveref}

\graphicspath{{equations/pictures/},{coupling_scheme/pictures/},{numerical_analysis/pictures/}}
\usetikzlibrary{positioning, shapes, arrows, calc, patterns}
\newcommand{\procor}{\texttt{PROCOR}}

\newcommand{\norm}[2][\null]{\left\Vert#2\right\Vert_{#1}}
\newcommand{\abs}[1]{\left\vert#1\right\vert}
\newcommand{\inner}[2]{\langle\,#1,\,#2\rangle}
\newcommand{\D}[3][\null]{\frac{\mathrm{d}^{#1}#2}{\mathrm{d}^{#1}#3}}

\newcommand{\domain}[2][\null]{\Omega_{#2}^{#1}}
\newcommand{\interface}[2][\null]{\Gamma_{#2}^{#1}}
\newcommand{\frontier}[2][\null]{\partial\Omega_{#2}^{#1}}
\newcommand{\Dt}{\Delta t}
\newcommand{\dt}{\delta t}

\newcommand{\dEnthalpy}[2][\null]{\Delta \mathcal{H}_{#2}^{#1}}

\newcommand{\heatCp}[2][\null]{C{\scriptstyle p}_{#2}^{#1}}
\newcommand{\heatCpUnits}{\text{J.kg$^{-1}$.K$^{-1}$}}
\newcommand{\mFlowRate}[2][\null]{\dot{m}_{#2}^{#1}}
\newcommand{\mFlowRateUnits}{\text{kg.s$ ^{-1}$}}

\newcommand{\massPower}[2][\null]{\dot{q}_{#2}^{#1}}
\newcommand{\massPowerUnits}{\text{W.kg$^{-1}$}}

\newcommand{\heatFlux}[2][\null]{\phi_{#2}^{#1}}

\newcommand{\heatFluxUnits}{\text{W.m$^{-2}$}}
\newcommand{\vol}[2][\null]{V_{#2}^{#1}}
\newcommand{\volUnits}{\text{m$^3$}}
\newcommand{\area}[2][\null]{S_{#2}^{#1}}
\newcommand{\areaUnits}{\text{m$^2$}}
\newcommand{\length}[2][\null]{e_{#2}^{#1}}
\newcommand{\lengthUnits}{\text{m}}
\newcommand{\mass}[2][\null]{m_{#2}^{#1}}
\newcommand{\massUnits}{\text{kg}}
\newcommand{\temperature}[2][\null]{T_{#2}^{#1}}
\newcommand{\temperatureUnits}{\text{K}}

\newcommand{\massDensity}[2][\null]{\rho_{#2}^{#1}}

\newcommand{\thermalCond}[2][\null]{\lambda_{#2}^{#1}}

\newcommand{\keywords}{Severe Accidents, Multiphysics, Coupling Scheme, Partitioned Approach, Stability Analysis, Lumped Parameter Model, Complex System, Computing Efficiency}
\newcommand{\lp}{LP}
\newcommand{\ecs}{ECS}
\newcommand{\ics}{ICS}
\hypersetup{
    bookmarks=true,         
    unicode=false,          
    pdftoolbar=true,        
    pdfmenubar=true,        
    pdffitwindow=false,     
    pdfstartview={FitH},    
    pdftitle={Solving coupled problems of lumped parameter models in a platform for severe accidents in nuclear reactors},
	pdfauthor={L. Viot, L. Saas, and F. De Vuyst},
	pdfkeywords={\keywords{}},
    pdfsubject={Subject},   
    pdfcreator={Creator},   
    pdfproducer={Producer}, 
    pdfnewwindow=true,      
    colorlinks=true,       
	allcolors=green!50!black,     
    urlcolor=red!50!black           
}
\title{Solving coupled problems of lumped parameter models in a platform for severe accidents in nuclear reactors}
\author{%
\large \textbf{Louis Viot$^{1,*}$, Laurent Saas $^2$ and Florian De Vuyst $^3$}\\%
$^1$\small CMLA, Ecole Normale Supérieure de Cachan, 94235 Cachan, France\\%
$^*$\small email : \href{mailto:louis.viot@cea.fr}{louis.viot@cea.fr}\\%
$^{1,2}$\small CEA, DEN, DTN/SMTA/LMAG, Cadarache, F-13108 Saint-Paul-lez-Durance, France\\%
$^3$\small LMAC, UTC, Sorbonne Universités, 60200 Compiègne, France%
}
\date{}

\begin{document}
\maketitle
\begin{abstract}
This paper focuses on solving coupled problems of lumped parameter models. Such problems are of interest for the simulation of severe accidents in nuclear reactors~: these coarse-grained models allow for fast calculations for statistical analysis used for risk assessment and solutions of large problems when considering the whole severe accident scenario. However, this modeling approach has several numerical flaws. Besides, in this industrial context, computational efficiency is of great importance leading to various numerical constraints. The objective of this research is to analyze the applicability of explicit coupling strategies to solve such coupled problems and to design implicit coupling schemes allowing stable and accurate computations. The proposed schemes are theoretically analyzed and tested within CEA's \procor{} platform on a problem of heat conduction solved with coupled lumped parameter models and coupled 1D models. Numerical results are discussed and allow us to emphasize the benefits of using the designed coupling schemes instead of the usual explicit coupling schemes. 
\end{abstract}
\begin{flushleft}
\end{flushleft}
\begin{flushleft}
\textbf{Acronyms. }\textit{\lp{}, lumped parameter; \ecs{}, explicit coupling scheme; \ics{}, implicit coupling scheme}
\end{flushleft}
\section{Introduction}
\label{sec:introduction}
Mathematical and numerical resolution of coupled multiscale and multiphysics problems is a substantial issue arising in several engineering fields. In particular in the field of severe accidents in nuclear reactors \cite{sehgal_nuclear_2012} \cite{jacquemain_les_2013} coupling thermohydraulics, thermomechanics, thermochemistry, thermodynamics, neutronics phenomena with characteristic time, length and mass going from microseconds to days, millimeters to meters, kilograms to hundred of tons. In shuch a context, this requires the simulation of the whole accident scenario or just a part of it leading to coupled problems of different size. Furthermore, simulation can be used for statistical analysis, e.g.\ Monte-Carlo sensitivity analysis, involving a large number of calculations, or simply to have a finer and better understanding of some phenomenons. Altogether and because of the lack of full physical and phenomenological knowledge, a wide range of models for the underlying physical phenomena are used, e.g. stationary models, reduced models, mesh based models, etc. In particular, \textbf{lumped parameter models} (\lp{} models), sometimes called ``0D'' models, simplifie spatially distributed systems into discrete entities, e.g.\ partial differential equations become parameterized ordinary differential equations over time, in such a way that calculations require much less running time and a much lower computational cost. However, such simplifications come with a price to pay:
\begin{itemize}
\item \lp{} models ignore the finite time of propagation of information of the continuous or space and time discretized model and instantly communicate and spread them; 
\item closure laws used in \lp{} models are often described with highly non-linear functions of time and of the model internal state variables, e.g.\ correlations fitted from mesh based calculations or experimental results \cite{bonnet_thermohydraulic_1999} \cite{zhang_natural_2015};
\item state, phase or topological changes in the continuous time and space model, e.g.\ disappearance, vaporization, are turned into instant internal changes and sometimes discontinuous event triggered at a certain time by the \lp{} model, which can be compared to event occuring in \emph{Differential Algebraic Equations} (DAE) \cite{mao_efficient_2002}. A state change event missed by the simulation engine can bring the coupled models in a non coherent and non physical state during a small window of time resulting in numerical errors, numerical instabilities which have to be avoided at all cost;
\end{itemize}
As a result, coupled problems of \lp{} models can have fast and stiff transients which are numerically challenging to solve. 

There are two approaches to solve such coupled problems : a \textbf{monolithic} approach which solves the governing equations describing each model simultaneously and a \textbf{partitioned} approach \cite{felippa_staggered_1980}, its counterpart, in which coupled models are associated to the so-called \emph{partitions} which are solved one at a time during the coupling iterations. One of the advantage of such partitioned approaches is that the solution of partitions can be done with different solvers adapted to the physical phenomenon involved in the partition. Moreover, great modularity and software reuse is achieved since partition solvers are assumed to be seen as ``black box'' by the partitioned problem with a set of inputs, a set of outputs and very limited internal details (e.g.\ derivative data) and thus can be easily exchanged. In this approach, partitions deliver physical quantities such as heat fluxes, forces, pressures, mass flow rates to other coupled partitions. In contrast with the monolithic approach, coupling equations between partitions are not part of a one block system of equations, instead, partitions are sharing data with external coupling equations corresponding to equilibrium conditions, e.g.\ heat flux equality between two thermal partitions sharing a geometrical interface \cite{giles_stability_1997} or temperature equality between a thermal partition and a thermodynamic partition sharing a common temperature. Because of possible decoupling effects between partitions, equilibrium conditions might not be enforced by the coupling algorithm leading to loosely or weakly coupled partitions. Therefore, there are two main classes of coupling schemes :
\begin{itemize}
\item \textbf{Explicit coupling schemes} (\ecs{}s) \cite{felippa_staggered_1980,farhat_mixed_1995,guillard_significance_2000,farhat_consistency_1990,farhat_robust_2010,farhat_unconditionally_1991} which require only one call of each solver one after the other during each time step but which can only achieve weak coupling between the coupled partitions at the end of the time step.
\item \textbf{Implicit coupling schemes} (\ics{}s) \cite{joosten_analysis_2009,michler_interface_2005,gerbeau_quasi-newton_2003,degroote_performance_2009,kassiotis_nonlinear_2011,vierendeels_implicit_2007,degroote_multi-level_2012,minami_performance_2010,ganine_nonlinear_2013,kuttler_fixed-point_2008} which require several calls of solvers within an iterative loop until a certain convergence threshold. If the scheme has converged, equations and subdomains are strongly coupled at the end of the time step, up to a certain precision, and the monolithic scheme is recovered.
\end{itemize}

In this paper, we focus on the numerical solution of heterogeneous 0D and 1D models by partitioned approaches. In this very specific context of modeling, we show how \ecs{}s are often not suitable for the solution of such problems and implicit treatment is often necessary. We furthermore explain how regular \ics{}s are modified in order to take into account state change events triggered by the models. By this way, the coupled models can be synchronized on potential events and discontinuities.\medskip

The paper is structured as follows. \Cref{sec:governing_equations} gives a brief overview of the lumped parameter mass and energy conservation governing equations in each subdomain and models used for the different previously cited phenomena ; for the sake of simplicity and conciseness, we only consider coupled thermalhydraulic phenomena. We believe that the extension to any other phenomena (e.g.\ thermochemistry, thermomechanic, neutronic, etc.) is similar. From there, the coupled formulation is explained and the coupling algorithms are given in \cref{sec:coupling_formulation}. In \cref{sec:numerical_analysis}, we give numerical analysis, computations and discussions of a coupled problem solved with various coupling schemes within the \procor{} platform \cite{le_tellier_phenomenological_2015}. Finally, conclusion and opening remarks are given in \cref{sec:conclusion}.

\section{Governing equations}
\label{sec:governing_equations}
\begin{figure}
    \centering
	\includegraphics[width=0.6\textwidth]{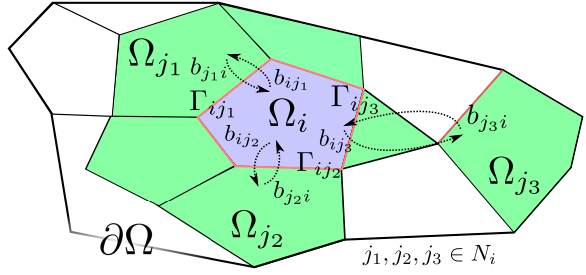}
	\caption{The abstract representation of domain decomposition of $\domain{}$ focused on domain $\domain{i}$ and its neighbourhood $\domain{j \in N_i}$.}
	\label{fig:domain_decomposition}
\end{figure}

To simplify, we assume that the coupled problem to solve consists of different physics located at different subdomains. \Cref{fig:domain_decomposition} depicts a non overlapping domain decomposition of domain $\domain{} = \cup_{i \in  \llbracket 1,m \rrbracket} \domain{i}$. Neighbouring domains $\domain{j}$ coupled with domain $\domain{i}$ communicating through interface $\interface{ij}$ can be actual geometrical neighbor of domain $\domain{i}$, i.e.\ $\interface{ij} = \frontier{i} \cap \frontier{j} \neq \emptyset$ or can be distant neighbor of domain $\domain{i}$ with whom it has linked but distant phenomena, i.e.\ $\frontier{i} \cap \frontier{j} = \emptyset$ and $\interface{ij}$ is a part of frontier $\frontier{i}$ of domain $\domain{i}$. Neighbors of domain $\domain{i}$ are represented by the set $N_i = \{j\;/\;\exists\;\interface{ij}\}$, with cardinality denoted by $n_i = \text{card}(N_i)$. Finally, frontier of domain $\domain{i}$ can be calculated by $\frontier{i} = \interface{i} \cup (\cup_{j \in N_i} \interface{ij})$.
Subdomain $\domain{i}$ is described in terms of mass denoted by $\mass{i}\,[\massUnits{}]$ and average temperature denoted by $\temperature{i}\,[\temperatureUnits{}]$. We have :\[\mass{i} = \frac{1}{V_{\scriptscriptstyle{\domain{i}}}}\int_{\domain{i}}\massDensity{}\,\mathrm{d}V,\quad\temperature{i} = \frac{1}{V_{\scriptscriptstyle{\domain{i}}}}\int_{\domain{i}}\temperature{}\,\mathrm{d}V.\] The vector of state variables of subdomain $\domain{i}$ is denoted by $\mathbf{u}_i = (m_i,T_i)^{\textrm{t}}$. Note that in this article we decide to represent each subdomain in term of average temperature even though they could be represented in term of average enthalpy.

Interface variables are heat fluxes $\heatFlux{ij}\,[\heatFluxUnits{}]$, temperature $\temperature{ij}\,[\temperatureUnits{}]$, mass flow rate $\mFlowRate{ij}\,[\mFlowRateUnits{}]$ or surface area $\area{ij}\,[\areaUnits{}]$. These variables are gathered in vector $\mathbf{b} = \{(\heatFlux{ij}, \temperature{ij}, \mFlowRate{ij}, \area{ij})^{\textrm{t}}\}$ for $i \in [1,m]$ and $j \in N_i$. The interface projector $\mathcal{P}_{ij}$ allows us to get interface $\interface{ij}$ vector variables $\mathbf{b}_{ij} = (\heatFlux{ij}, \temperature{ij}, \mFlowRate{ij}, \area{ij})^{\textrm{t}}$ from vector $\mathbf{b}$ with $\mathcal{P}_{ij} \mathbf{b} = \mathbf{b}_{ij}$.

We first describe the lumped parameter governing equations for each subdomain in \cref{sec:subdomain_equations} and then present the interface equations between subdomains in \cref{sec:interface_equations}.
\subsection{Subdomain lumped parameter equations}
\label{sec:subdomain_equations}
Subdomain $\domain{i}$ equations are expressed in terms of mass and energy macroscopic conservation equations (momentum conservation equations are modeled by closure laws, e.g. correlations for heat transfer coefficient). They are obtained from the local conservation equations, here Navier-Stokes equations under the Boussinesq approximation for liquid domains and heat equations for solid domains, integrated over the corresponding subdomain (see \cite{le_tellier_treatment_2017}). Tightly linked to the local physical model, this approach leads to the so-called \lp{} model or ``0D'' model of the subdomain  described by the two ordinary differential equations (ODE)
\begin{align}
\D{\mass{i}}{t}  &= \sum_{j \in N_i} \mFlowRate{ij}\quad\text{in}\;\domain{i} \label{eq:0D_mass_equation},\\
\mass{i} \heatCp{i} \D{\temperature{i}}{t} + \sum_{j \in N_i} \mFlowRate{ij} \heatCp{i} (\temperature{i} - \temperature{ij})&= \sigma_i \heatFlux{i} \area{i} + \sum_{j \in N_i} \sigma_{ij} \heatFlux{ij} \area{ij} + \mass{i} \massPower{i}\quad\text{in}\;\domain{i} \label{eq:0D_temperature_equation}
\end{align}
with $\mass{i}\,[\massUnits{}]$ the mass and $\temperature{i}\,[\temperatureUnits{}]$ the average temperature of $\domain{i}$, $\heatFlux{i}\,[\heatFluxUnits{}]$ the heating ($\sigma_{i} = 1$) or cooling ($\sigma_{i} = -1$) heat flux through boundary $\interface{i} = \frontier{} \cap \frontier{i}$ with temperature $\temperature{b_i}$ and area $\area{i}\,[\areaUnits{}]$, $\heatFlux{ij}$ the heating or cooling heat flux and $\mFlowRate{ij}\,[\mFlowRateUnits{}]$ the algebraic mass flow rate through $\interface{ij}$ with temperature $\temperature{ij}$ and area $\area{ij}$. Finally, $\heatCp{i}\,[\heatCpUnits{}]$ is the heat capacity and $\massPower{i}\,[\massPowerUnits{}]$ is the residual power per mass unit coming from fission products of subdomain $\domain{i}$.

The previous physical parameters are obtained from closure laws described hereafter in \cref{sec:interface_equations}. In particular, geometry dependent values, i.e.\ the characteristic length of a domain or the surface of an interface or the volume of a domain, are given by algebraic geometry equations. They take the form of algebraic functions, e.g.\ for the previous surface : 
\begin{equation}
\label{eq:0D_geometry}
\area{ij} = \area{ij}(\massDensity{i},\length{i},\vol{i})
\end{equation}
with $\length{i}\,[\lengthUnits{}]$ the characteristic length and $\vol{i}\,[\volUnits{}]$ the volume of domain $\domain{i}$. Altogether, ordinary differential \cref{eq:0D_mass_equation,eq:0D_temperature_equation} combined with expressions like \cref{eq:0D_geometry} can be considered as \textit{Differential Algebraic Equations} (DAE's) \cite{petzold_differential/algebraic_1982} describing the \lp{} model.

\subsection{Interface equations}
\label{sec:interface_equations}
In the present article, interfaces between two subdomains $\domain{i}$ and $\domain{j}$ can be of two types : free moving or fixed boundaries. Physically moving boundaries are, for example, boundaries between a liquid and a solid domain exchanging melted or solid materials. They are associated with a plane fusion solidification front corresponding to the Stefan condition at the interface (see \cite{le_tellier_treatment_2017} for further details). Fixed interfaces correspond to thermal equilibrium assuming no mass exchange through the interface and thermal conduction.  

In case of a mobile interface, equilibrium conditions at the interface $\interface{ij}$ are given by
\begin{align}
\label{eq:fusion_front_equilibrium_condition_1}
\mFlowRate{ij} &= -\mFlowRate{ji}\quad\text{in}\;\interface{ij},\\
\label{eq:fusion_front_equilibrium_condition_2}
\heatFlux{ij}\area{ij} &= - \heatFlux{ji}\area{ji} + \dEnthalpy[\textrm{fus.}]{} \mFlowRate{ij}\quad\text{in}\;\interface{ij},\\
\label{eq:fusion_front_equilibrium_condition_3}
\temperature{ij} &= \temperature{ji} = \temperature[\textrm{fus.}]{}\quad\text{in}\;\interface{ij}
\end{align}
with $\dEnthalpy[\textrm{fus.}]{}$ the fusion enthalpy and $\temperature[\textrm{fus.}]{}$ the fusion temperature of domain $\domain{i}$, both assumed to be fixed. In particular, those conditions stipulate that the mass flow rate should be the same on both sides of the interface for mass conservation and that the heat fluxes should respect the Stefan condition. To simplify subdomain materials are treated as pure body and no thermochemistry is considered. 

In case of a fixed interface, the thermal equilibrium conditions at interface $\interface{ij}$ are given by
\begin{align}
\label{eq:thermal_equilibrium_condition_1}
\mFlowRate{ij} &= -\mFlowRate{ji} = 0\quad\text{in}\;\interface{ij},\\
\label{eq:thermal_equilibrium_condition_2}
\temperature{ij} &= \temperature{ji}\quad\text{in}\;\interface{ij},\\
\label{eq:thermal_equilibrium_condition_3}
\heatFlux{ij}\area{ij} &= -\heatFlux{ji}\area{ji}\quad\text{in}\;\interface{ij}
\end{align}

Obviously, with the previous equations, appropriate closure laws for interface heat fluxes and temperatures are required. In traditional mesh based models, interface variables are given by a projection operator which are coherent with the domain equations, e.g.\ the restriction of the variables values over all the domain $\domain{i}$ to its interface $\interface{ij} = \frontier{i} \cap \frontier{j}$, and thus are consistent with the physical equations. For \lp{} models, interface variable $\mathbf{b}_{ij}$ is calculated from spatially averaged data $\mathbf{u}_i$ from domain $\domain{i}$ and data $(\{\mathbf{b}_{ij}\}_{j \in N_i})  \overset{\textit{def}}{=} \mathbf{b}_{i\star}$ from the interfaces. They are given by closure laws which take the form of algebraic functions $\mathbf{b}_{ij} = \mathbf{b}_{ij}(\mathbf{u}_i, \mathbf{b}_{i\star})$. Thus they propagate data instantly, from one specific interface to all the other ones. 

If the domain $\domain{i}$ is solid, such closure law functions for heat fluxes $\heatFlux{i\bullet}$ can be calculated from the heat diffusion conduction equation under certain assumptions and approximations. A comparison of those different approximate models with a reference solution given by a finite element discretization of the heat conduction equation can be found in \cite{le_tellier_treatment_2017}. For example, the \textit{stationary} model uses a 1D cylindrical with adiabatic lateral boundaries approximation and assumes a quadratic temperature profile in the solid domain. Conduction heat fluxes at interfaces $\interface{ij}$ and $\interface{ik}$ associated with the upper and lower cylinder surfaces are then given by the temperature derivative w.r.t the spatial direction leading to the closure law functions \cite{le_tellier_treatment_2017} :
\begin{align}
\heatFlux{ij} &= \heatFlux{ij}(\mathbf{u}_i, \mathbf{b}_{ik}) = \thermalCond{i} \frac{6 \temperature{i} - 4 \temperature{ij} - 2 \temperature{ik}}{\length{i}}, \label{eq:heat_flux_stationary_model_1}\\
\heatFlux{ik} &= \heatFlux{ik}(\mathbf{u}_i, \mathbf{b}_{ij}) = \thermalCond{i} \frac{6 \temperature{i} - 4 \temperature{ik} - 2 \temperature{ij}}{\length{i}}. \label{eq:heat_flux_stationary_model_2}
\end{align}
Note, for instance, the instant propagation of interface $\interface{ik}$ temperature $\temperature{ik}$ to interface $\interface{ij}$ in \cref{eq:heat_flux_stationary_model_1} indicating that the \textit{stationary} model gives closure law functions which propagate boundary related data \textbf{instantly} between interfaces of the domain.

Let us mention that, in the literature, one can find other nonlinear closure laws like $\heatFlux{ij}(\mathbf{u}_i, \mathbf{b}_{i\star}) \propto \length[\alpha]{i} (\temperature{i}-\temperature{ij})^{\beta}$ with $\alpha,\beta \in \mathbb{R}$ for convective heat transfer or $\heatFlux{ij}(\mathbf{u}_i, \mathbf{b}_{i\star}) \propto {\temperature{ij}}^4$ for radiative heat transfer. In this paper, we will only consider \cref{eq:heat_flux_stationary_model_1,eq:heat_flux_stationary_model_2}.

\section{Coupling formulation and schemes}
\label{sec:coupling_formulation}
\subsection{Discretized coupling formulation and coupled problem}
\lp{} conservation equations \cref{eq:0D_mass_equation,eq:0D_temperature_equation} are then discretized in time at a lower level with a method of choice (e.g.\ explicit, implicit, Euler, Runge-Kutta, multistep methods, etc.) and then coupled in time on a higher level resulting in a two-level time scheme. At the highest level, those discretized equations are solved and coupled between times $t^0,t^1,\ \ldots,t^n,\ \ldots$ with a macro time step~$\Dt$ and synchronized at each of these times. At the lowest level, each subdomain manages its own time integration scheme, its own micro time step $\dt$ and its own internal time. The integration scheme used by the subdomain is assumed to be adapted to the physical local problem. In the following, discretized values evaluated at time $t^n$ are denoted by the superscript $n$.

We adopt the following notations. For each coupled subdomain $\{ \domain{i} \}_{i\in[1,m]}$, discretized equations are represented by a function $\mathcal{F}_i^{\Dt}$ used to solve and advance in time the problem of one time step $\Dt$. This function takes as parameters the state vector $\mathbf{u}_i$ and the input interface variables $\{\mathbf{b}_{ji}\}_{j \in N_i} \overset{\text{def.}}{=} \mathbf{b}_{\star i}$ of domain $\domain{i}$. Equations of the coupled problem are then given by 
\begin{equation}
\label{eq:flow_formulation}
\left\{
	\begin{aligned}
		\mathcal{F}_1^{\Dt}(\mathbf{u_1},&\ \{\mathbf{b}_{j1}(\mathbf{u}_j,\ \mathbf{b}_{\star j})\}_{j \in N_1}) = 0\\
		\mathcal{F}_2^{\Dt}(\mathbf{u_2},&\ \{\mathbf{b}_{j2}(\mathbf{u}_j,\ \mathbf{b}_{\star j})\}_{j \in N_2}) = 0\\
		&\vdots\\
		\mathcal{F}_m^{\Dt}(\mathbf{u_m},&\ \{\mathbf{b}_{jm}(\mathbf{u}_j,\ \mathbf{b}_{\star j})\}_{j \in N_m}) = 0\\
	\end{aligned}
\right.
\end{equation}
in such a way that the inter dependencies between domain $\domain{i}$ and its neighbors are highlighted by the closure law functions $\{\mathbf{b}_{ji}\}_{j \in N_i}$ which is function of the coupled subdomain internal variable $\mathbf{u}_j$ and interface variables $\mathbf{b}_{\star j}$, which might in turn need values from domain $\domain{i}$ to be calculated. 

Those interface dependencies highlight the input-output relationship between interface variables of the different subdomains. This suggests that solvers can be seen as closed entities or ``black-boxes'' taking input interface variables $\mathbf{b}_{\star i}$ from neighboring subdomains and giving back output interface variables $\mathbf{b}_{i\star}$ to neighboring subdomains. Therefore, it seems natural to represent solver of domain $\domain{i}$ as a function $\mathcal{M}_i^{\Dt} : \mathbb{R}^{3 \times n_i} \mapsto \mathbb{R}^{3 \times n_i}$ in which we explicitly hide subdomain state variable $\mathbf{u}_i$ and its integration, and only the input and output interface variables are visible. It is possible that we have access to limited information about them, e.g.\ no derivative data. For instance, solver of domain $\domain{i}$ is defined by
\begin{equation}
\label{eq:solver_formulation}
\mathbf{b}_{i\star} = \mathcal{M}_i^{\Dt}(\mathbf{b}_{\star i}).
\end{equation}

From \cref{eq:solver_formulation}, we can define the coupled problem at interface $\interface{ij}$ in term of solvers :
\begin{align}
\mathbf{b}_{ij} &= \mathcal{P}_{ij} \circ \mathcal{M}_i^{\Dt}\left(\mathbf{b}_{ji},\{\mathbf{b}_{ki}\}_{k \in N_i,\ k \neq j}\right)\label{eq:ij_projection},\\
\mathbf{b}_{ji} &= \mathcal{P}_{ji} \circ \mathcal{M}_j^{\Dt}\left(\mathbf{b}_{ij},\{\mathbf{b}_{kj}\}_{k \in N_j,\ k \neq i}\right)\label{eq:ji_projection}
\end{align}
with $\mathcal{P}_{ij}$ the interface $\interface{ij}$ projector. Those equations emphasize the ``action-reaction'' at interface between the two domains : a slight modification of interface variable $\mathbf{b}_{ji}$ will in turn change the associated interface variable $\mathbf{b}_{ij}$, and vice versa. The strength of the associated coupling is represented by the Jacobian matrices $\{\frac{\partial \mathcal{M}_i^{\Dt}}{\partial \mathbf{b}_{ki}}\}_{k \in N_i}$ and $\{\frac{\partial \mathcal{M}_j^{\Dt}}{\partial \mathbf{b}_{kj}}\}_{k \in N_j}$. Moreover it is difficult to evaluate this strength since they may not be available.

Combining \cref{eq:ij_projection} with \cref{eq:ji_projection} gives the widely used fixed point equation at interface $\interface{ij}$ (see \cite{mehl_parallel_2016})
\begin{equation}
\label{eq:fpe_formulation}
\mathbf{b}_{ij} = \mathcal{P}_{ij} \circ \mathcal{M}_i^{\Dt}\left(\mathcal{P}_{ji} \circ \mathcal{M}_j^{\Dt}\left(\mathbf{b}_{ij},\{\mathbf{b}_{kj}\}_{k \in N_j,\ k \neq i}\right),\{\mathbf{b}_{ki}\}_{k \in N_i,\ k \neq j}\right)
\end{equation}
which allows us to define the residual operator $\mathcal{R}_{ij}$ at interface $\interface{ij}$  by
\begin{equation}
\label{eq:residual_formulation}
\mathcal{R}_{ij}(\mathbf{b}_{ij}) \overset{\textit{def}}{=} \mathcal{P}_{ij} \circ \mathcal{M}_i^{\Dt}\left(\mathcal{P}_{ji} \circ \mathcal{M}_j^{\Dt}\left(\mathbf{b}_{ij},\{\mathbf{b}_{kj}\}_{k \in N_j,\ k \neq i}\right),\{\mathbf{b}_{ki}\}_{k \in N_i,\ k \neq j}\right) - \mathbf{b}_{ij}
\end{equation}
for a guess candidate $\mathbf{b}_{ij}$. The residual $\mathcal{R}_{ij}(\mathbf{b}_{ij})$ will express the imbalance created at interface $\interface{ij}$ : if domains $\domain{i}$ and $\domain{j}$ are strongly coupled, equilibrium conditions at interface are fulfilled and the interface residual is null, otherwise the two domains are only weakly coupled.

\subsection{Coupling schemes}
\label{sec:coupling_scheme}
\paragraph{Explicit coupling schemes (\ecs{}s).}Referred to as ``conventional serial staggered'' in \cite{felippa_staggered_1980}, \ecs{}s solve the coupled problem with one call of solver per time step. They are based on a Gauss-Seidel semi-explicit solution of the coupled problem \cref{eq:flow_formulation}. Obviously, another scheme based on a fully-explicit or Jacobi solution can be used to allow for more algorithm parallelism. If we assume that solver $\mathcal{M}_i^{\Dt}$ is solved before solver $\mathcal{M}_j^{\Dt}$, the scheme solves at interface $\interface{ij}$ :
\begin{align}
\mathbf{b}_{ij}^{n+1} &= \mathcal{P}_{ij} \circ \mathcal{M}_i^{\Dt}\left(\mathbf{b}_{ji}^{n},\{\mathbf{b}_{ki}^{\bullet}\}_{k \in N_i,\ k \neq j}\right),\\
\mathbf{b}_{ji}^{n+1} &= \mathcal{P}_{ji} \circ \mathcal{M}_j^{\Dt}\left(\mathbf{b}_{ij}^{n+1},\{\mathbf{b}_{kj}^{\bullet}\}_{k \in N_j,\ k \neq i}\right)
\end{align}
with $\mathbf{b}_{ki}^{\bullet}$ is evaluated at $t^n$ or $t^{n+1}$ if solver $\mathcal{M}_k^{\Dt}$ is called before or after solver $\mathcal{M}_i^{\Dt}$.

While potentially attractive and fast since only one call per solver is done during each time step, it is well known that this scheme yields poor accuracy and stability issues \cite{piperno_partitioned_1995}. Because of the \textit{time-lag} caused by the semi-explicit resolution, it is unlikely that interface residuals defined by \cref{eq:residual_formulation} are null and equilibrium conditions are not enforced at interface. Besides, despite several improvements and studies in \cite{farhat_mixed_1995} \cite{guillard_significance_2000} \cite{farhat_consistency_1990} \cite{farhat_robust_2010} \cite{farhat_unconditionally_1991} \cite{piperno_partitioned_2001}, the weak coupling reached by explicit coupling schemes is often not enough and only a strong coupling at the end of the time step can ensure proper stability properties \cite{causin_added-mass_2005} \cite{giles_stability_1997}. 

\paragraph{Implicit coupling schemes (\ics{}s).} A way to fix the \textit{time-lag} is to use implicit coupling between subdomains. At interface $\interface{ij}$, implicit coupling gives~:
\begin{align}
\mathbf{b}_{ij}^{n+1} &= \mathcal{P}_{ij} \circ \mathcal{M}_i^{\Dt}\left(\mathbf{b}_{ji}^{n+1},\{\mathbf{b}_{ki}^{n+1}\}_{k \in N_i,\ k \neq j}\right)\label{eq:implicit_coupling_1},\\
\mathbf{b}_{ji}^{n+1} &= \mathcal{P}_{ji} \circ \mathcal{M}_j^{\Dt}\left(\mathbf{b}_{ij}^{n+1},\{\mathbf{b}_{kj}^{n+1}\}_{k \in N_j,\ k \neq i}\right)\label{eq:implicit_coupling_2}
\end{align}

We clearly see that \cref{eq:implicit_coupling_1,eq:implicit_coupling_2} give discrete interface values respecting the fixed point \cref{eq:fpe_formulation} leading to a null interface residual. Hence, interface $\interface{ij}$ is at equilibrium and a strong coupling between equations is obtained. However, while being mathematically interesting, \cref{eq:implicit_coupling_1,eq:implicit_coupling_2} do not allow to decouple the two equations at each time step and it is often more convenient to use iterative methods to solve them. By iterative methods we mean methods using iterative processes inside one coupling iteration, i.e. between times $t^n$ and $t^{n+1}$, until a convergence criterion is reached. In that case interface variables \textit{verify} (with an error bounded by the convergence criterion) at each interface the associated fixed point \cref{eq:fpe_formulation} leading to a strong coupling between equations of the coupled problem. At interface $\interface{ij}$ between domain $\domain{i}$ and $\domain{j}$, the block Gauss-Seidel coupling scheme, also known as staggered or partitioned coupling scheme, is an iterative method based on Newton-Raphson iterations of the fixed point \cref{eq:fpe_formulation} given by 
\begin{equation}
\label{eq:newton_iterations}
\mathbf{b}_{ji}^{n+1, k+1}  = \mathbf{b}_{ji}^{n+1, k}  + \left(\frac{\mathrm{d}\mathcal{R}_{ji}}{\mathrm{d}\mathbf{b}_{ji}}\Bigr|_{\mathbf{b}_{ji}^{n+1, k}}\right)^{-1} (-\mathcal{R}_{ji}(\mathbf{b}_{ji}^{n+1,k}))
\end{equation}
with the residual $\mathcal{R}_{ji}(\mathbf{b}_{ji}^{n+1, k})$ at iteration $k$ defined by
\begin{equation}
\label{eq:residual}
\mathcal{R}_{ji}(\mathbf{b}_{ji}^{n+1, k}) = \mathbf{r}_{ji}^{n+1,k} = \mathcal{M}_j^{\Dt} \circ \mathcal{M}_i^{\Dt}(\mathbf{b}_{ji}^{n+1, k}) - \mathbf{b}_{ji}^{n+1, k} 
\end{equation}
which allows us to define a convergence criterion for the iterative process by
\begin{equation}
\label{eq:convergence}
\frac{\norm{\mathcal{M}_j^{\Dt} \circ \mathcal{M}_i^{\Dt}(\mathbf{b}_{ji}^{n+1, k}) - \mathbf{b}_{ji}^{n+1, k}}}{\norm{\mathbf{b}_{ji}^{n+1, k}}} = \frac{\norm{\mathbf{r}_{ji}^{n+1,k}}}{\norm{\mathbf{b}_{ji}^{n+1, k}}} \leq \epsilon_{\textrm{rel}}
\end{equation}
with $\epsilon_{\textrm{rel}}$ the relative stopping criterion for the subiterative process. \textit{Gauss-Seidel} or \textit{staggered} symbolizes the way interfaces data are sequenced through solvers (or \textit{block}) inside coupling iterations. 
\paragraph{Remark on Jacobian matrices and associated solvers.} In \cref{eq:newton_iterations}, the Jacobian matrix or its inverse are usually not known and/or not calculable since solvers are seen as black-box solvers, thus the iterations can only be approximated by approximation of the Jacobian matrix or its inverse leading to Quasi-Newton techniques. In \cite{gerbeau_quasi-newton_2003}, Gerbeau and al. use reduced order models to calculate the Jacobian matrix. In \cite{michler_interface_2005}, Michlet and al. use Newton-Krylov subiterations to approximate the Jacobian matrix leading to the so-called \textit{Jacobian free Newton-Krylov} method. From previous residuals, Degroote and al. in \cite{degroote_performance_2009} solve least-squares problems to approximate the inverse of the Jacobian matrix leading to \textit{Interface Quasi-Newton - Inverse Least-Squares} (IQN-ILS) methods while Vierendeels and al. \cite{vierendeels_implicit_2007} approximate the Jacobian matrix with similar methods leading to \textit{Interface Quasi-Newton - Least-Squares} (IQN-LS). Multigrid Quasi-Newton methods are also studied in \cite{degroote_multi-level_2012}. A recent review of different Quasi-Newton methods can be found in \cite{minami_performance_2010} for fluid-structure interaction and in \cite{ganine_nonlinear_2013} for thermal-structure interaction.

Much cheaper methods use relaxation techniques for the interface variable iterations which consists in using an approximation of the inverse of the interface residual Jacobian matrix of the form
\begin{equation}
\left(\frac{\mathrm{d}\mathcal{R}_{ji}}{\mathrm{d}\mathbf{b}_{ji}}\Bigr|_{\mathbf{b}_{ji}^{n+1, k}}\right)^{-1} \approx -\omega^k\,I
\end{equation}
for which the fixed point iterations with dynamic relaxation are given by
\begin{equation}
\label{eq:relaxation}
\mathbf{b}_{ji}^{n+1,k+1} = \omega^k \mathcal{M}_j^{\Dt} \circ \mathcal{M}_i^{\Dt}(\mathbf{b}_{ji}^{n+1, k}) + (1-\omega^k) \mathbf{b}_{ji}^{n+1, k}.
\end{equation}
Note that for $\omega^k=1$, iterations given by \cref{eq:relaxation} are classical Picard iterations which converge generally only linearly and very slowly. In \cite{kuttler_fixed-point_2008}, the authors show that the block Gauss-Seidel iterative method with relaxation techniques for the resolution of equation \cref{eq:newton_iterations} can be very efficient at a surprisingly low cost in comparison to more elaborated Quasi-Newton methods. Besides, it is shown in \cite{ramiere_iterative_2015} that these techniques are also proven to be very competitive in comparison to the direct solution of the nonlinear problem given by \cref{eq:newton_iterations} with traditional Newton-Raphson methods using derivative data. In this paper, we focus on block Gauss-Seidel iterative method with relaxation techniques. 

For instance, the solution of equation \cref{eq:newton_iterations} by the Steffensen's method accelerated with Aitken's delta-squared $\Delta^2$ method \cite{aitken_xxv.bernoullis_1927} can be cast into fixed point iterations with dynamic relaxation leading to methods of order $2$, see \cite{ramiere_iterative_2015}. Less costly but with a convergence rate of the golden ratio $\frac{1+\sqrt{5}}{2}$, the fixed point iterations given by the secant method can also be seen as fixed point iterations with dynamic relaxation leading to
\begin{equation}
\omega^k = -\omega^{k-1}\frac{\inner{\mathbf{r}_{ji}^{n+1,k}-\mathbf{r}_{ji}^{n+1,k-1}}{\mathbf{r}_{ji}^{n+1,k-1}}}{\inner{\mathbf{r}_{ji}^{n+1,k}-\mathbf{r}_{ji}^{n+1,k-1}}{\mathbf{r}_{ji}^{n+1,k}-\mathbf{r}_{ji}^{n+1,k-1}}}.
\end{equation}
One can also use constant relaxation given by a constant parameter $\omega$. However, this requires the determination of the best relaxation parameters, i.e. leading to the highest rates of convergence, which is highly problem-dependent.

\begin{figure}
    \centering
	\includegraphics[width=0.7\textwidth]{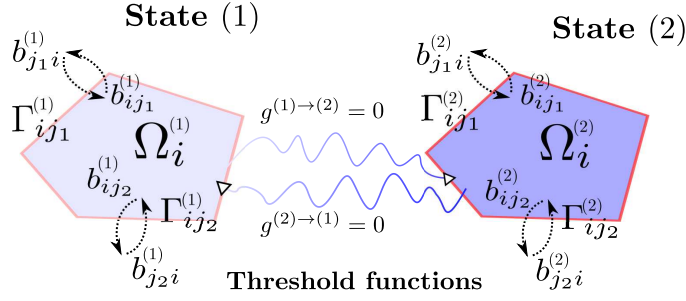}
	\caption{The abstract representation of states $\domain[(1)]{i}$ and $\domain[(2)]{i}$ of domain $\domain{i}$. Each state has its own interfaces $\interface[(1)]{\bullet}$ and $\interface[(2)]{\bullet}$ with associated variables $\mathbf{b}^{(1)}_{\bullet}$ and $\mathbf{b}^{(2)}_{\bullet}$. A state transition occurs when a transition function crosses zero.}
	\label{fig:domain_states}
\end{figure}

\paragraph{Implicit coupling schemes for event detection and model synchronization.} In \lp{} modeling, state transitions are triggered by internal events corresponding to activation of threshold functions. This activation is dependent on input from the coupled models and thus times of events are unknowns of the coupled problem. As depicted \cref{fig:domain_states}, each state has its own solver, its own set of equations and its own interface and boundary conditions. 

State transition can lead to discontinuities of state or interface variables. A missed event can lead to inconsistent and incoherent physical states, e.g. the disappearance of a model not seen by other coupled models, and/or mathematical difficulties with non-smooth functions in the coupled problem \cref{eq:flow_formulation} bringing usual theorems out of their scope of validity, with possibly breakdown issue like divergence of Newton iterations. 

During one coupling time step $\Dt$, the coupling scheme must be able to adapt its time step to synchronize all the models on the time of the first event. However, synchronization strategies are hard to achieve with \ecs{}s which can often only synchronized coupled models at each coupled time step leading to time detection errors of order $\mathcal{O}(\Dt)$. Hereafter, we describe how the previous \ics{}s can be used for event detection to ensure proper synchronization between models.\medskip

\begin{figure}
\begin{tikzpicture}
\node at (-2.3,0.) {$\mathcal{M}_i$ \textbf{internal} time loop};
\draw [->, thick] (0,0) --  (8.3,0);
\node at (-2.3,-1.0) {$\mathcal{M}_j$  \textbf{internal} time loop};
\draw [->, thick] (0,-1) --  (8.3,-1);
\node at (-2.3,-2.0) {\textbf{Shared} time loop};
\draw [->, thick] (0,-2) --  (8.3,-2);				
\draw[thick] (0.1,-0.1) -- (0.1,0.1);
\draw[thick] (0.1,-1.1) -- (0.1,-0.9);
\draw[thick] (0.1,-2.1) -- (0.1,-1.9);
\node at (0.1,-2.5) {$t^n$};
\draw[thick] (8.0,-0.1) -- (8.0,0.1);
\draw[thick] (8.0,-1.1) -- (8.0,-0.9);
\draw[thick] (8.0,-2.1) -- (8.0,-1.9);
\node at (8.2,-2.5) {$t^n+\Dt$};
\node at (6.8,-1.6) {$\textcolor{blue!50!black}{t^{\star,k}}$};
\draw (6.7,-1.9) -- (6.9,-2.1);
\draw (6.7,-2.1) -- (6.9,-1.9);
\node at (6,0.4) {$\textcolor{red!50!black}{t_i^{\star}}$};
\draw (5.9,0.1) -- (6.1,-0.1);
\draw (5.9,-0.1) -- (6.1,0.1);
\node at (5.2,-0.6) {$\textcolor{red!50!black}{t_j^{\star}}$};
\draw (5.1,-0.9) -- (5.3,-1.1);
\draw (5.1,-1.1) -- (5.3,-0.9);
\node at (5.2,-1.6) {$\textcolor{red!50!black}{t^{\star}}$};
\draw (5.1,-1.9) -- (5.3,-2.1);
\draw (5.1,-2.1) -- (5.3,-1.9);
\node at (3.0,0.4) {$\textcolor{blue!50!black}{t_i^{\star,k}}$};
\draw (2.9,0.1) -- (3.1,-0.1);
\draw (2.9,-0.1) -- (3.1,0.1);
\node at (4.2,-0.6) {$\textcolor{blue!50!black}{t_j^{\star,k}}$};
\draw (4.1,-0.9) -- (4.3,-1.1);
\draw (4.1,-1.1) -- (4.3,-0.9);
\node at (2.0,-1.6) {$\textcolor{blue!50!black}{\tilde{t}^{\star,k}}$};
\draw (1.9,-1.9) -- (2.1,-2.1);
\draw (1.9,-2.1) -- (2.1,-1.9);
\node at (3.5,-1.6) {$\textcolor{blue!50!black}{t^{\star,k+1}}$};
\draw (3.4,-1.9) -- (3.6,-2.1);
\draw (3.4,-2.1) -- (3.6,-1.9);
\node at (3.5,-2.4) {$\scriptstyle\alpha \textcolor{blue!50!black}{\tilde{t}^{\star,k}} + (1-\alpha)\textcolor{blue!50!black}{t^{\star,k}}$};
\end{tikzpicture}
\caption{Shared coupling time loop and internal time loops of solvers $\mathcal{M}_i$ and $\mathcal{M}_j$. Internal events at times $\textcolor{red!50!black}{t_i^{\star}}$ and $\textcolor{red!50!black}{t_j^{\star}}$ are to be detected by the scheme in order to synchronize on time $\textcolor{red!50!black}{t^{\star}}$ of the first event. An \ics{} detects event at times $\textcolor{blue!50!black}{t_i^{\star,k}}$ and $\textcolor{blue!50!black}{t_j^{\star,k}}$ and has to calculate the new iteration $\textcolor{blue!50!black}{t^{\star,k+1}}$ in such a way that $\textcolor{blue!50!black}{t^{\star,\infty}} \approx \textcolor{red!50!black}{t^{\star}}$.}
\label{fig:events_time_loop}
\end{figure}
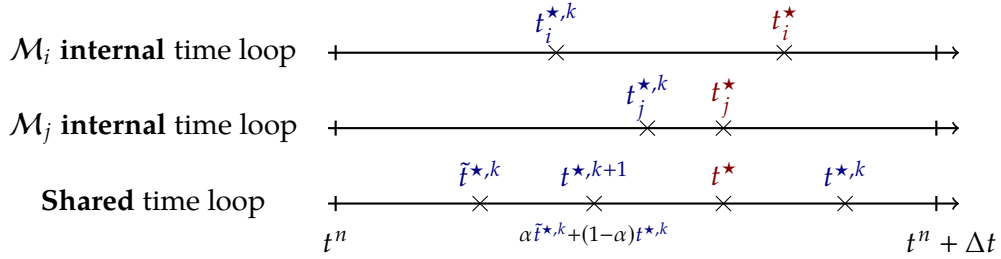

Let us consider the coupling at interface $\interface{ij}$ between domain $i$ and $j$. As depicted in \cref{fig:events_time_loop}, internal events are triggered during one coupling time step $\Dt$ by solver $\mathcal{M}_i$ and/or solver $\mathcal{M}_j$ at respective times $\textcolor{red!50!black}{t_i^{\star}}$ and $\textcolor{red!50!black}{t_j^{\star}}$ between $t^n$ and $t^n+\Dt$. When an event occurs, the solver stops its computation and do not compute the whole coupling time step $\Dt$. Solvers should be synchronized on the first triggered event at time $\textcolor{red!50!black}{t^{\star}} = \textcolor{red!50!black}{t_j^{\star}}$.

During a fixed point iteration of an \ics{} starting at time $t^n$ and ending at the previously calculated time of first event $\textcolor{blue!50!black}{t^{\star,k}}$, events are trigerred at times $\textcolor{blue!50!black}{t_i^{\star,k}}$ and $\textcolor{blue!50!black}{t_j^{\star,k}}$ by solvers. The aim is to build an iterative algorithm calculating the new iteration $\textcolor{blue!50!black}{t^{\star,k+1}}$ in such a way that~: $\textcolor{blue!50!black}{t^{\star,\infty}} \approx \textcolor{red!50!black}{t^{\star}}$. \Cref{alg:event_time_iteration} gives a sequence to calculate this new iteration.
\begin{algorithm}
\caption{Calculate the new iteration $\textcolor{blue!50!black}{t^{\star,k+1}}$}
\label{alg:event_time_iteration}
\renewcommand{\algorithmiccomment}[1]{/* #1 */}
\algsetup{indent=2em}
\begin{algorithmic}[1]
\STATE Compute a new fixed point iteration between solvers $\mathcal{M}_i^{\Dt}$ and $\mathcal{M}_j^{\Dt}$ with an \ics{};
\IF {\textit{an event has occurred for model $l\in\{i,j\}$}} 
        \STATE $\textcolor{blue!50!black}{t_l^{\star,k}}\gets$ the corresponding event time;
\ELSE
        \STATE $\textcolor{blue!50!black}{t_l^{\star,k}} \gets t^n+\Dt$;
		\STATE \COMMENT{the whole coupling time step $\Dt$ was computed.}
\ENDIF
\STATE $\textcolor{blue!50!black}{\tilde{t}^{\star,k}} \gets \min{}(\textcolor{blue!50!black}{t_i^{\star,k}},\textcolor{blue!50!black}{t_j^{\star,k}});$
\IF[$\epsilon_{\textrm{rel}}$ being a tolerance given by the \ics{}.] {$(\vert\textcolor{blue!50!black}{\tilde{t}^{\star,k}} - \textcolor{blue!50!black}{t^{\star,k}}\vert / \Dt < \epsilon_{\textrm{rel}}$) \AND (\textit{the \ics{} has converged at interface} $\interface{ij}$)} 
	\STATE $\textcolor{blue!50!black}{t^{\star,k+1}} \gets \textcolor{blue!50!black}{\tilde{t}^{\star,k}};$
	\STATE $\textcolor{blue!50!black}{t^{\star,\infty}} \gets \textcolor{blue!50!black}{t^{\star,k+1}};$
\ELSE
	\STATE $\textcolor{blue!50!black}{t^{\star,k+1}} \gets \alpha \textcolor{blue!50!black}{\tilde{t}^{\star,k}} + (1-\alpha)\textcolor{blue!50!black}{t^{\star,k}};$
	\STATE \COMMENT{$\alpha\in]0,1[$ being a relaxation parameter given by the \ics{}.}
\ENDIF
\end{algorithmic}
\end{algorithm} 
At convergence, the coupled problem has been solved between times $t^n$ and $t^{n+1} = \textcolor{blue!50!black}{t^{\star,\infty}}$. The next coupling time step then starts at time $t^{n+1}$ with the same algorithm.
\paragraph{Concluding remarks regarding the designed iterative algorithm.}At convergence $k=\infty$ of a coupling time step between times $t^n$ and $t^{n+1}$ at interface $\interface{ij}$, we get that :
\begin{itemize}
\item Model $i$ and model $j$ are strongly coupled because the fixed point \cref{eq:fpe_formulation} is verified with at most an error bounded by the tolerance $\epsilon_{\textrm{rel}}$ of the scheme, i.e. \[\frac{\norm{\mathcal{M}_j^{\Dt} \circ \mathcal{M}_i^{\Dt}(\mathbf{b}_{ji}^{n+1, \infty}) - \mathbf{b}_{ji}^{n+1, \infty}}}{\norm{\mathbf{b}_{ji}^{n+1, \infty}}} \leq \epsilon_{\textrm{rel}}.\]
\item Model $i$ and model $j$ are synchronized on potential internal triggered events. In particular, they are synchronized on the time of the first event with an error also bounded by the tolerance $\epsilon_{\textrm{rel}}$, i.e. \[\frac{\abs{t^{n+1} - t^{\star}}}{\Dt} = \frac{\abs{t^{\star, \infty} - t^{\star}}}{\Dt} \leq  \epsilon_{\textrm{rel}}\]
\end{itemize}

\section{Numerical analysis and examples}
\label{sec:numerical_analysis}
\label{sec:numerical_results}
In the following, the aim is to present numerical solutions of some coupled \lp{} models implemented in the CEA's \procor{} platform with both \ecs{}s and \ics{}s. The industrial \procor{} platform \cite{le_tellier_phenomenological_2015} allows generic ``black-box'' physical models coupling solved by various \ecs{}s and \ics{}s. It is dedicated to the fast robust setup of coupled problems taking the form of complex systems \cite{boccara_modeling_2010} for the simulation of severe accidents in nuclear reactors. For computational efficiency reasons in this industrial context, one important constraint is that we want to be able to keep a sufficiently large coupling time step in comparison to the characteristic times of the physical phenomena being involved. With this constraint, it will be shown that even ``simple'' coupled problems of \lp{} models may produce unexpected artifacts in terms of coupling and synchronization and how the use of \ics{}s to solve them can provide benefits. 

\begin{figure}
    \centering
	\begin{subfigure}[b]{.7\textwidth}
        \centering
		\begin{tikzpicture}[node distance=0cm]
		\filldraw[draw=black, line width=0.1mm, shade, top color=red!90, bottom color=red!50]  plot[smooth, tension=.7] coordinates {(-4,0.) (-3,0.9) (-2,0.6) (-1,0.8) (0,0.9) (1,0.85) (2,0.9) (2.5,0.7) (3,0)} -- node[pos=0.7, yshift=0.3cm] {$\scriptstyle\domain{1} :\ \mass{1},\temperature{1},\massDensity{1},\heatCp{1},\thermalCond{1}$} (-4,0);
		\filldraw[draw=black, line width=0.1mm, shade, top color=red!50, bottom color=red!10]  plot[smooth, tension=.4] coordinates {(-4,0) (-3,-0.9) (-2,-0.75) (-1,-0.7) (0,-0.8) (1,-0.92) (2,-0.75) (2.5,-0.7) (3,0)} -- node[pos=0.7, yshift=-0.3cm] {$\scriptstyle\domain{2} :\ \mass{2},\temperature{2},\massDensity{2},\heatCp{2},\thermalCond{2}$} (-4,0);
		\draw [->, thick] (-4.2, -1.1) -- node [pos=1, left] {$\scriptstyle z$} (-4.2,1.1);
		\draw [draw=black] (-4.3, -1.08) -- node [left] {$\scriptstyle 0$} (-4.1,-1.08);
		\draw [draw=black, dotted] (-4.3, 0) -- (0,0);
		\draw [draw=black, dotted] (-4.3, 0.92) -- (3.,0.92);
		\draw [draw=black, dotted] (-4.3, -0.92) -- (3.,-0.92);
		\node at (-4.5,-0.75) {$\scriptstyle\length{2}$};
		\node at (-4.5,0.75) {$\scriptstyle\length{1}$};
		\node at (-4.5,0.) {$\interface{12}$};
		\draw [->, double] (0.5,0.5) -- (0.5,0.1) node [midway,right] {$\scriptstyle\heatFlux{12},\ \mFlowRate{12},\ \temperature{12}$};
		\draw [->, double] (0.5,-0.5) -- (0.5,-0.1) node [midway,right] {$\scriptstyle\heatFlux{21},\ \mFlowRate{21},\ \temperature{21}$};
		\node at (-0.5,1.2) {$\scriptstyle\temperature{b_1} \gg \temperature{b_2}\ \text{in}\ \frontier{1} \backslash \interface{12}$};
		\node at (-0.5,-1.2) {$\scriptstyle\temperature{b_2} \ll \temperature{b_1}\ \text{in}\ \frontier{2} \backslash \interface{12}$};
		\end{tikzpicture}
		\label{fig:conduction_notations}
    \end{subfigure}%
	\begin{subfigure}[b]{.3\textwidth}
        \begin{tikzpicture}[node distance=1.5cm]
		\tikzstyle{vertex}=[ellipse, thick, align=center, draw=black, minimum height=1.0cm]
		\tikzstyle{transition}=[->, thick, bend right=10]
		\tikzstyle{state}=[rectangle, thick, draw=black, minimum height = 1.0cm, minimum width=2.0cm, align=center]
		\node[vertex, shade, top color=red!90, bottom color=red!50] (1) {$\scriptstyle \heatFlux{12} = \mathcal{M}_1(\temperature{21},\ \mFlowRate{21})$};
		\node[vertex, shade, top color=red!50, bottom color=red!10, below of= 1] (2) {$[\scriptstyle \temperature{21},\ \mFlowRate{21}] = \mathcal{M}_2(\heatFlux{12})$};
		\path 	(1) edge[transition] node[pos=0.5, left] {$\scriptstyle \heatFlux{12}$} (2)
		(2) edge[transition] node[pos=0.5, right] {$\scriptstyle [\temperature{21},\ \mFlowRate{21}]$} (1);	
		\end{tikzpicture}
		\label{fig:conduction_coupling}
    \end{subfigure}
	\caption{Heat conduction between domains $\domain{1}$ and $\domain{2}$ : notations (\textit{left}) and Dirichlet-Neumann coupling of solvers $\mathcal{M}_1$ and $\mathcal{M}_2$ (\textit{right}).}
	\label{fig:toy_example}
\end{figure}
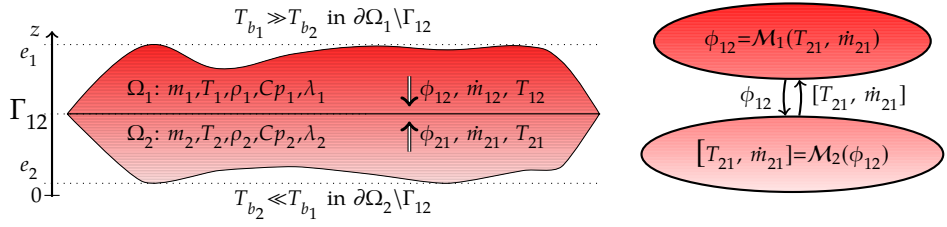

For instance, let us consider the heat conduction between domains $\domain{1}$ and $\domain{2}$ calculated respectively by solvers $\mathcal{M}_1 $ and $\mathcal{M}_2$. As depicted in \cref{fig:toy_example}, solvers are coupled at interface $\interface{12} = \frontier{1} \cap \frontier{2}$ of unit area $\area{}=1$ with Dirichlet-Neumann boundary conditions given by \cref{eq:fusion_front_equilibrium_condition_1,eq:fusion_front_equilibrium_condition_2,eq:fusion_front_equilibrium_condition_3} or \cref{eq:thermal_equilibrium_condition_1,eq:thermal_equilibrium_condition_2,eq:thermal_equilibrium_condition_3}. This coupling ensure the well posedness of the domain decomposition problem \cite{dolean_introduction_2015}. For this coupled problem, two types of heat conduction solvers are used : \lp{} solvers and finer solvers based on a space discretization of the 1D heat equation. Finally, we assume a cylindrical geometry in such a way that the length and mass of each domain are related with $\mass{\{1,2\}} = \massDensity{\{1,2\}} \area{} \length{\{1,2\}} = \massDensity{\{1,2\}} \length{\{1,2\}}$.

\subsection{Linear stability of the coupling of lumped parameter solvers}
\label{sec:discontinuity_numerical_analysis}
First, we consider the Dirichlet-Neumann boundary conditions at the fixed interface $\interface{12}$ given by \cref{eq:thermal_equilibrium_condition_1,eq:thermal_equilibrium_condition_2,eq:thermal_equilibrium_condition_3} $\mFlowRate{12} = -\mFlowRate{21} = 0, \temperature{12} = \temperature{21}, \heatFlux{12} = -\heatFlux{21}$. The two domain masses are constant and only their energy conservation equations are calculated. In the following, we analyze the linear stability of the coupling of \lp{} solvers. They are based on a time discretization of the energy conservation \cref{eq:0D_temperature_equation} and closure laws \cref{eq:heat_flux_stationary_model_1,eq:heat_flux_stationary_model_2} to solve the heat conduction problem. A similar stability analysis for finer 1D solvers can be found in \cite{giles_stability_1997}.
\subsubsection{Stability of a \textit{toy} explicit coupling scheme}
\label{sec:toy_explicit_coupling}
The aim here is to build a prototype of a `toy' \ecs{} with no subcycling and only one level of time discretization. Such coupling scheme should allow us to solve the coupled problem while solvers of domains $\domain{1}$ and $\domain{2}$ are only called once during a time step. To do so, we use an implicit Euler scheme for the first domain $\domain{1}$ and an explicit Euler scheme for the second domain $\domain{2}$ . Denoting by $\Delta t$ the time step, the discretized equations for the first domain $\domain{1}$ are given by
\begin{equation}
\label{eq:explicit_coupling_scheme_1}
\left\{
	\begin{aligned}
		\heatFlux[n]{12} &= \thermalCond{1} \frac{6 \temperature[n]{1} - 2 \temperature{b_1} - 4 \temperature[n]{12}}{\length{1}}\quad\text{in}\;\domain{1}\\
		\massDensity{1} \heatCp{1} \length{1} \frac{\temperature[n]{1} - \temperature[n-1]{1}}{\Delta t} &= -\heatFlux[n]{12}\quad\text{in}\;\domain{1}
	\end{aligned}
\right.
\end{equation}
with continuity of temperature at the interface, i.e.
\begin{equation}
\label{eq:explicit_coupling_scheme_12}
	\temperature[n]{12} = \temperature[n]{21}\quad\text{on}\;\interface{12}
\end{equation}
and for the second domain $\domain{2}$
\begin{equation}
\label{eq:explicit_coupling_scheme_2}
\left\{
	\begin{aligned}
		\temperature[n+1]{21} &= -\frac{1}{4}\frac{\length{2}}{\thermalCond{2}} \heatFlux[n]{21} + \frac{3}{2} \temperature[n+1]{2} - \frac{1}{2} \temperature{b_2}\quad\text{in}\;\domain{2}\\
		\massDensity{2} \heatCp{2} \length{2} \frac{\temperature[n+1]{2} - \temperature[n]{2}}{\Delta t} &= -\heatFlux[n]{21}\quad\text{in}\;\domain{2}\\
	\end{aligned}
\right.
\end{equation}
with continuity of heat flux at the interface, i.e.
\begin{equation}
\label{eq:explicit_coupling_scheme_21}
		\heatFlux[n]{21} = -\heatFlux[n]{12}\quad\text{on}\;\interface{12}.
\end{equation}
The \ecs{} is sequenced as follows : given an interface temperature $\temperature[n]{21}$ at time $t^n$, the first domain is advanced from time $t^{n-1}$ to $t^n$ and a heat flux $\heatFlux[n]{12}$ is computed at time $t^n$ which is then imposed to the second domain. It is then advanced from time $t^n$ to $t^{n+1}$ and it computes a new temperature $\temperature[n+1]{21}$ at time $t^{n+1}$, and so on. Thus, the coupling scheme only requires one call to each solver per time step so that the scheme can be called explicit.

We now analyze the linear stability of the coupling scheme. Combining \cref{eq:explicit_coupling_scheme_1,eq:explicit_coupling_scheme_12,eq:explicit_coupling_scheme_2,eq:explicit_coupling_scheme_21}, we get
\begin{equation}
\label{eq:linear_difference_equation}
	\left(1+6\frac{\Delta t}{\tau_1}\right)\heatFlux[n+1]{12} - \left(1+\left(1-6\frac{\Delta t}{\tau_2}\right)\hbar\right)\heatFlux[n]{12} + \hbar\,\heatFlux[n-1]{12} = 0
\end{equation}
with $\hbar=\frac{\thermalCond{1} / \length{1}}{\thermalCond{2} / \length{2}}$, the characteristic times of conduction $\tau_1 = \massDensity{1} \heatCp{1} {\length{1}}^2 / \thermalCond{1}$ in domain $\domain{1}$ and $\tau_2 = \massDensity{2} \heatCp{2} {\length{2}}^2 / \thermalCond{2}$ in domain $\domain{2}$. To ensure stability of the explicit scheme in domain $\domain{2}$, $\Delta t$ has to be small in comparison to $\tau_2$, i.e.\ $\Delta t / \tau_2 \ll 1$. The simplified characteristic polynomial $\chi$ associated to the simplified second order linear difference \cref{eq:linear_difference_equation} has two complex roots $x_{\{1,2\}}^{\star}$ which have to be of modulus strictly smaller than one in order to guaranty a stable coupling scheme. 

We get the following results :
\begin{itemize}
\item When $\Delta t / \tau_1 \ll 1$ and $\hbar<1$, the two roots are real and a Taylor expansion gives 
\begin{align*}
	x_1^{\star} &= 1-\frac{6}{1-\hbar}\frac{\Delta t}{\tau_1} + \mathcal{O}(\left(\frac{\Delta t}{\tau_1}\right)^2)\\
	x_2^{\star} &= \hbar+\frac{6\hbar^2}{1-\hbar}\frac{\Delta t}{\tau_1} + \mathcal{O}(\left(\frac{\Delta t}{\tau_1}\right)^2)
\end{align*}
When $\Delta t / \tau_1 \ll 1$ and $\hbar>1$, the two roots are complex of modulus \[\abs{x_{\{1,2\}}^{\star}}^2 = \frac{\hbar}{1+6\frac{\Delta t}{\tau_1}}\]
\item For any values of $\Delta t / \tau_1$, when $\hbar \ll 1$, roots of the characteristic polynomial $\chi$ are $x_1^{\star} \approx \frac{1}{1+6\frac{\Delta t}{\tau_1}}$ and $x_2^{\star} \approx \frac{\hbar}{1+6\frac{\Delta t}{\tau_1}}$ and when $\hbar \gg 1$,  $x_1^{\star} \approx 1+\frac{6\frac{\Delta t}{\tau_1}}{\hbar}$ and $x_2^{\star} \approx \frac{\hbar}{1+6\frac{\Delta t}{\tau_1}}$.
\end{itemize}

Thus, in both cases when $\hbar>1$ the spurious solution for $\heatFlux{12}$ of \cref{eq:linear_difference_equation} associated to root $x_2^{\star}$ increases with a growth rate of $\frac{\hbar}{1+6\,\Delta t / \tau_1}$. In most cases, the spurious solution will grow in time leading to an unstable scheme. On the contrary, cases when $\hbar<1$ give a stable coupling scheme.

Thus, the remaining question is the value $\hbar^{\textrm{crit}}$ of $\hbar$ above which the instabilities appear, thus defining the stability limit region of the \ecs{}. Further calculations show that $\hbar^{\textrm{crit}} = \abs{\frac{1+\ 6\ \Delta t/\tau_1}{1-\ 6\ \Delta t / \tau_2}}$ allowing to define a \textit{pseudo} CFL condition for the linear stability of the \ecs{} :
\begin{equation}
\label{eq:CFL_explicit}
r_{12}(\Delta t) \overset{\textit{\tiny def.}}{=} \abs{\frac{1-\ 6\ \Delta t / \tau_2}{1+\ 6\ \Delta t/\tau_1}} \hbar < 1
\end{equation}

\begin{figure}
    \centering
	\begin{subfigure}[b]{.5\textwidth}
        \includegraphics[width=\textwidth]{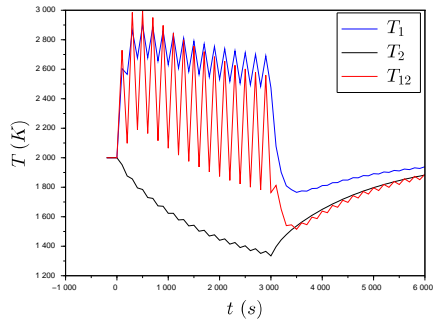}
		\caption{\lp{} solvers, $\hbar=1.6$}
		\label{fig:explicit_thermal_interface_11}
    \end{subfigure}%
	\begin{subfigure}[b]{.5\textwidth}
        \includegraphics[width=\textwidth]{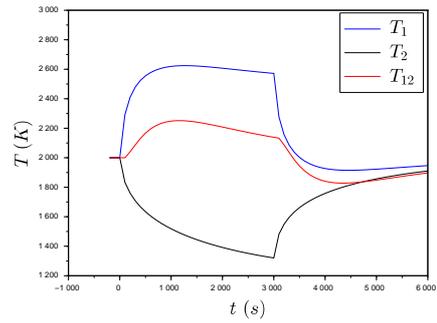}
		\caption{multidimensional solvers, $\hbar=1.6$}
		\label{fig:explicit_thermal_interface_12}
    \end{subfigure}\\
	\begin{subfigure}[b]{.5\textwidth}
        \includegraphics[width=\textwidth]{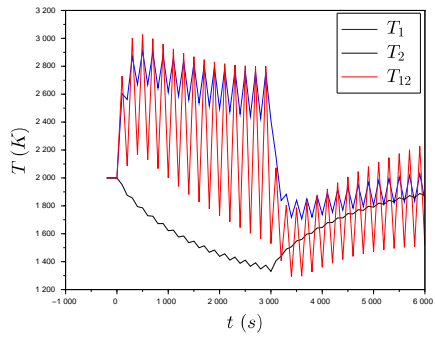}
		\caption{$\hbar=1.63$}
		\label{fig:explicit_thermal_interface_21}
    \end{subfigure}%
	\begin{subfigure}[b]{.5\textwidth}
        \includegraphics[width=\textwidth]{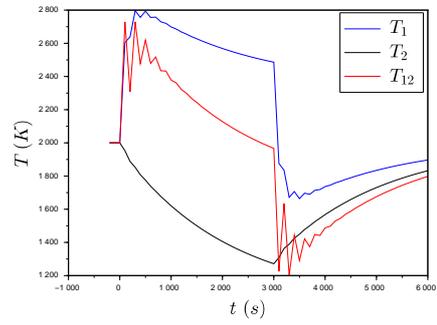}
		\caption{$\hbar=1.0$}
		\label{fig:explicit_thermal_interface_22}
    \end{subfigure}
	\caption{Propagation of discontinuities at boundaries in the explicit coupling of lumper parameter and finer (upper right) solvers for heat conduction for different values of $\hbar$ with $\Delta t / \tau_2 = 1/100$ and $\Delta t / \tau_1 = 1/10$.}
	\label{fig:explicit_thermal_interface}
\end{figure}

Finally, the Dirichlet-Neumann explicit coupling should be done in such a way that the domain with the Dirichlet boundary condition, i.e.\ imposed boundary temperature, has the lower thermal conductivity or the higher characteristic length and the domain with the Neumann boundary condition, i.e.\ imposed boundary heat flux, has the higher thermal conductivity and the lower characteristic length. Thus, the stability limit region $\hbar^{\textrm{crit}}$ can be extended for larger values of $\Delta t / \tau_1$ which can be explained by the diffusive properties of the \ics{} used in domain $\domain{1}$.
\paragraph{Numerical evidence of stability.} This analysis is illustrated by \cref{fig:explicit_thermal_interface}. At $t=0^-$, both domains are at an equilibrium temperature of $\temperature{1} = \temperature{2} = \temperature{21} = \temperature{b_1} = \temperature{b_2} = 2000$ K. Boundary discontinuities are imposed at $t=0^+$, $\temperature{b_1} = 3000$ K and $\temperature{b_2} = 400$ K and at $t=3\,\tau_1$, $\temperature{b_1} = \temperature{b_2} = 2000$ K. In addition, all of the computations use the value $\Delta t / \tau_2 = 1/100$ and $\Delta t / \tau_1 = 1/10$ corresponding to a stability limit region $\hbar^{\textrm{crit}} = 1.6$. \Cref{fig:explicit_thermal_interface_11} shows that when $\hbar = \hbar^{\textrm{crit}}$ the oscillations created by the transient initiated by the boundary discontinuities are not damped but are stable while in contrary in \cref{fig:explicit_thermal_interface_21} with $\hbar < \hbar^{\textrm{crit}}$ the oscillations are well damped and in \cref{fig:explicit_thermal_interface_22} with $\hbar > \hbar^{\textrm{crit}}$ the oscillations are clearly unstable. However, those three figures show that the discontinuities at boundaries are instantly propagated to the interface temperature and to the coupled domain in only one coupling time step. For comparison purpose, solution given by the coupling of the 1D heat equation solvers is given in \cref{fig:explicit_thermal_interface_12}. For the computation, the same physical values as for the coupling of \lp{} solvers are used. The values meet the requirements expressed in \cite{giles_stability_1997} to ensure proper stability of such solvers. For this coupling, the discontinuities at boundaries start a much slower and smaller transient ($\temperature{21}(t=1000\,\text{s}) \approx 2150$ K) than for the \lp{} coupling ($\temperature{21}(t=100\,\text{s}) \approx 2700$ K). The fact that \lp{} models propagate in one coupling time step all their discontinuities to the other models tend to create fast and important transient. Thus, the use of \ecs{} might not be adapted for this case. 

\subsubsection{Stability of a \textit{toy} implicit coupling scheme}
The iterative algorithm is described below. Given an initial boundary temperature $\temperature[n+1,0]{21}$ for domain $\domain{1}$, it iterates over $k \geq 0$ with the following steps~:
\begin{enumerate}
\item Use an \textbf{implicit} solver in $\domain{1}$ to find the interface heat flux $\heatFlux[n+1,k+1]{12}$
\begin{equation}
\label{eq:implicit_coupling_scheme_1}
\left\{
	\begin{aligned}
		\heatFlux[n+1,k+1]{12} &= \thermalCond{1} \frac{6 \temperature[n+1]{1} - 2 \temperature{b_1} - 4 \temperature[n+1]{12}}{\length{1}}\quad\text{in}\;\domain{1}\\
		\massDensity{1} \heatCp{1} \length{1} \frac{\temperature[n+1]{1} - \temperature[n]{1}}{\Delta t} &= -\heatFlux[n+1,k+1]{12}\quad\text{in}\;\domain{1}
	\end{aligned}
\right.
\end{equation}
with continuity of temperature at the interface, i.e.
\begin{equation}
\label{eq:implicit_coupling_scheme_12}
	\temperature[n+1]{12} = \temperature[n+1,k]{21}\quad\text{on}\;\interface{12}.
\end{equation}
\item Use an \textbf{implicit} solver in $\domain{2}$ to find the interface temperature $\tilde{T}^{n+1,k+1}_{21}$%
\begin{equation}
\label{eq:implicit_coupling_scheme_2}
\left\{
	\begin{aligned}
		 \tilde{T}^{n+1,k+1}_{21} &= -\frac{1}{4}\frac{\length{2}}{\thermalCond{2}} \heatFlux[n+1]{21} + \frac{3}{2} \temperature[n+1]{2} - \frac{1}{2} \temperature{b_2}\quad\text{in}\;\domain{2}\\
		 \massDensity{2} \heatCp{2} \length{2} \frac{\temperature[n+1]{2} - \temperature[n]{2}}{\Delta t} &= -\heatFlux[n+1]{21}\quad\text{in}\;\domain{2}\\
	\end{aligned}
\right.
\end{equation}
with continuity of heat flux at the interface, i.e.
\begin{equation}
\label{eq:implicit_coupling_scheme_21}
		\heatFlux[n+1]{21} = -\heatFlux[n+1,k+1]{12}\quad\text{on}\;\interface{12}.
\end{equation}
\item Measure convergence level with
\begin{equation}
	\frac{\abs{\tilde{T}^{n+1,k+1}_{21} - \temperature[n+1,k]{21}}}{\temperature[n+1,k]{21}} \leq \epsilon_{\textrm{rel}}.
\end{equation} 
If the previous predicate is not satisfied, relax the interface temperature by
\begin{equation}
	\temperature[n+1,k+1]{21} = \omega \tilde{T}^{n+1,k+1}_{21} + (1-\omega)\temperature[n+1,k]{21}
\end{equation}
and loop again. Otherwise, the final interface heat flux and temperature are given by $\temperature[n+1]{21} = \temperature[n+1,k]{21}$ and $\heatFlux[n+1]{12} = \heatFlux[n+1,k]{12}$. Consequently, internal variables $\temperature[n+1]{1}$ and $\temperature[n+1]{2}$ can be fully implicitly calculated : the iterative algorithm effectively allows to use implicit solvers for the two domains while allowing to decouple the two domains. 
\end{enumerate}

Combining \cref{eq:implicit_coupling_scheme_1,eq:implicit_coupling_scheme_12,eq:implicit_coupling_scheme_2,eq:implicit_coupling_scheme_21}, simple calculations lead to the following equation giving the interface temperature iterates $\temperature[n+1,k]{21}$ 
\begin{equation}
\temperature[n+1,k+1]{21} = \left[1-\left(1+\frac{1-6\frac{\Delta t}{\tau_2}}{1+6\frac{\Delta t}{\tau_1}}\hbar\right)\omega\right]\temperature[n+1,k]{21} + g(\temperature[n]{1}, \temperature[n]{2}, \temperature{b_1}, \temperature{b_2}).
\end{equation}
We note $\temperature[n+1]{21}$ the solution of the fixed point equation associated with the previous equation and $e_k =\abs{\temperature[n+1,k]{21} - \temperature[n+1]{21}}$ the error at iteration $k$. Then iteration errors are given by equation
\begin{equation}
\label{eq:error_toy_implicit_coupling}
e_{k+1} = \abs{1-\left(1+\frac{1-6\frac{\Delta t}{\tau_2}}{1+6\frac{\Delta t}{\tau_1}}\hbar\right)\omega} e_k.
\end{equation}
Thus, the iterative algorithm converges toward the interface solutions, i.e.\ $\temperature[n+1,\infty]{21} = \temperature[n+1]{21}$ and $\heatFlux[n+1, \infty]{12} = \heatFlux[n+1]{12}$, if and only if
\begin{equation}
\label{eq:omega_condition_convergence}
0 < \omega < \frac{2}{1+\frac{\abs{1-6\frac{\Delta t}{\tau_2}}}{\abs{1+6\frac{\Delta t}{\tau_1}}}\hbar} = \frac{2}{1+r_{12}(\Delta t)}.
\end{equation}
Note that when $\frac{\Delta t}{\tau_1} \ll 1$ and  $\frac{\Delta t}{\tau_2} \ll 1$, condition \cref{eq:omega_condition_convergence} reads 
\begin{equation}
	0 < \omega < \frac{2}{1+\hbar}.
\end{equation}
In cases where $\hbar>1$ for which the \ecs{} diverges, the iterative algorithm needs $\omega < 1$, i.e. under-relaxation to converge, which might lead to slow convergence.

\begin{figure}
    \centering
	\begin{subfigure}[b]{.5\textwidth}
        \includegraphics[width=\textwidth]{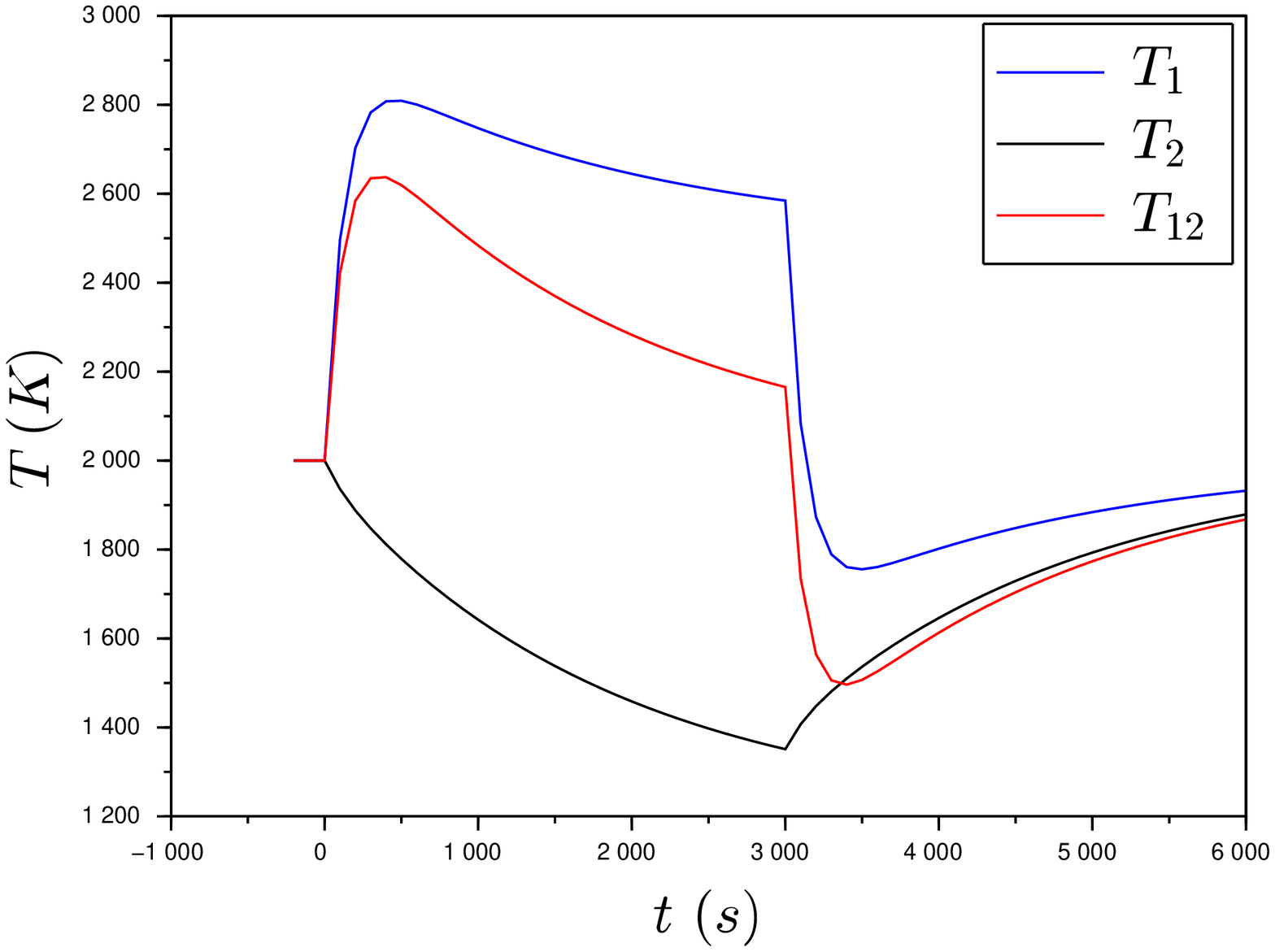}
		\caption{\ics{}, $\hbar=1.6$}
		\label{fig:implicit_thermal_interface_11}
    \end{subfigure}%
	\begin{subfigure}[b]{.5\textwidth}
        \includegraphics[width=\textwidth]{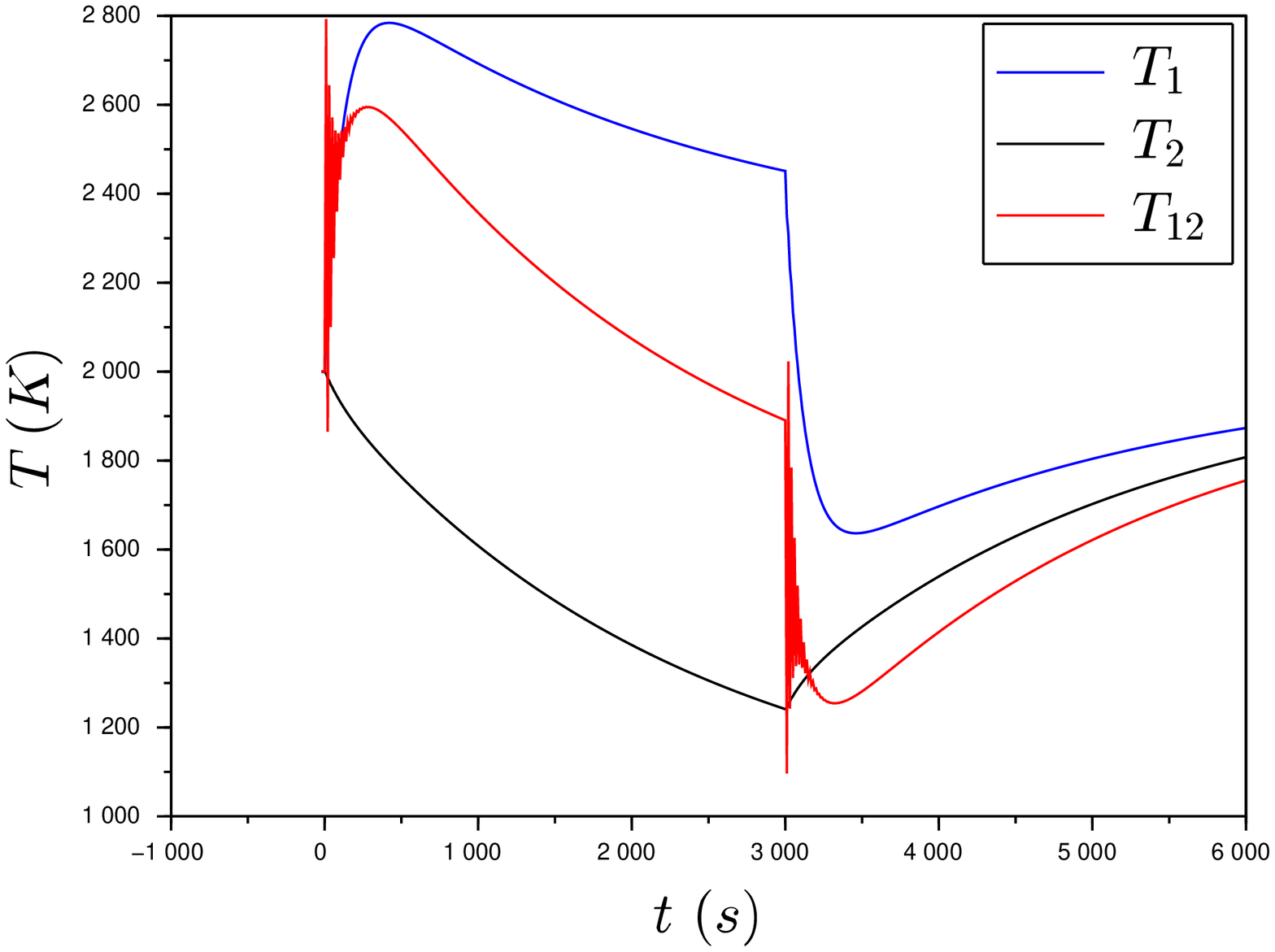}
		\caption{\ecs{}, $\hbar=0.625$, $\Delta t \leftarrow \Delta t / 10$}
		\label{fig:implicit_thermal_interface_12}
    \end{subfigure}\\
	\begin{subfigure}[b]{.5\textwidth}
        \includegraphics[width=\textwidth]{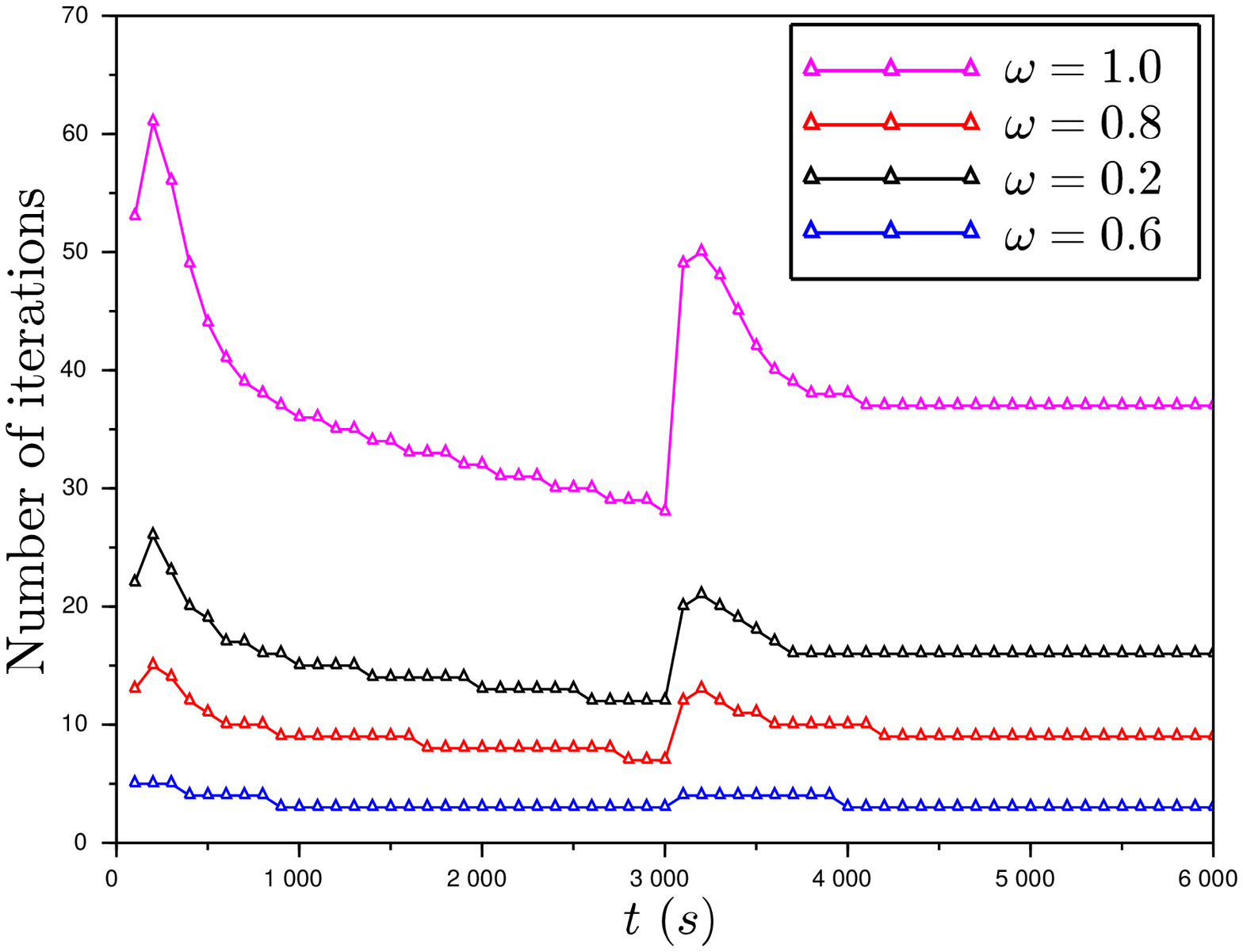}
		\caption{\ics{} number of iterations, $\hbar=1.6$}
		\label{fig:implicit_thermal_interface_21}
    \end{subfigure}%
	\begin{subfigure}[b]{.5\textwidth}
        \includegraphics[width=\textwidth]{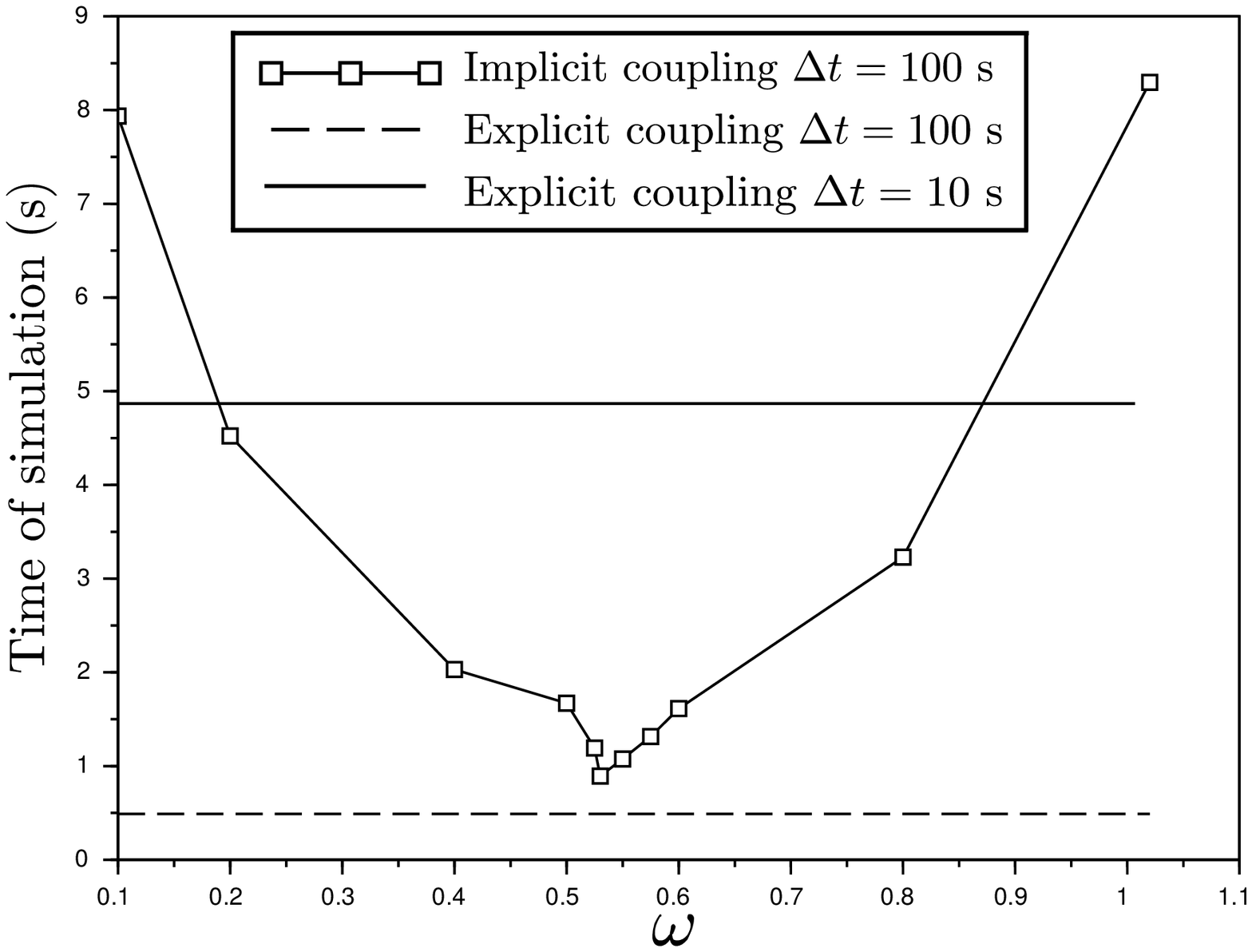}
		\caption{Schemes time of simulation, $\hbar=1.6$}
		\label{fig:implicit_thermal_interface_22}
    \end{subfigure}
	\caption{Propagation of discontinuities at boundaries of lumper parameter heat conduction coupled solvers solved with an \ics{} (upper left) and an \ecs{} (upper right) with much smaller coupling time step.}
	\label{fig:implicit_thermal_interface}
\end{figure}

\paragraph{Numerical evidence of stability.} This analysis is confirmed by \cref{fig:implicit_thermal_interface}. As in \cref{sec:toy_explicit_coupling}, both domains are at an equilibrium temperature of $2000$ K and the same discontinuities at boundaries are occurring at the same time $t=0^{+}$ and $t=3\,\tau_1$. For the computations, we still use $\Delta t / \tau_2 = 1/100$, $\Delta t / \tau_1 = 1/10$. We use $\hbar=1.6$ corresponding to the stability limit region for the \ecs{} ($\hbar = \hbar^{\textrm{crit}} = 1.6$) for which this scheme is showing constant oscillations. For this value, the corresponding maximal value of $\omega$ to ensure convergence of the iterative process is $\omega \approx 1.03$. The relative tolerance used is $\epsilon_{\textrm{rel}}=10^{-4}$. Several computations confirm the predicted behavior of the \ics{} : as long as $\omega$ stays under the calculated maximum value $1.03$, the \ics{} converges toward the same smooth solution given by \cref{fig:implicit_thermal_interface_11}. Even for values $\hbar > \hbar^{\textrm{crit}}$, the \ics{} handles smoothly the fast dynamics and important transient initiated by the instant propagation of the discontinuities at boundaries in the two domains by the \lp{} models. For cases where the \ecs{} is not unconditionally unstable, i.e. $r_{12}(\Delta t) < 1$, one way to achieve stability is to reduce the coupling time step : \cref{fig:implicit_thermal_interface_12} shows an explicit coupling solution with $\hbar = 0.625 < \hbar^{\textrm{crit}} \approx 1$ and coupling time step $\Delta t$ reduced by a factor of $10$. However, while effectively reducing the oscillations, using an \ecs{} with such a small coupling time step leads to higher computational time. The \ics{} allows to keep a sufficiently large coupling time step and a limited number of iterations as shown in \cref{fig:implicit_thermal_interface_21,fig:implicit_thermal_interface_22}. In the last figure, it is important to understand that only the solution given by the \ecs{} with a small coupling time step is usable but is costly to obtain while the solution given with the \ics{} is good and costs $5$ times less. For this problem, the optimal relaxation parameter leading to the higher convergence rate can be analytically calculated from \cref{eq:error_toy_implicit_coupling} and is given by $\omega = 1/(1+r_{12}(\Delta t)) \approx 0.52$. However, most of the time the optimal relaxation parameter is highly dependent to the problem.

\subsection{Synchronization on internal events}
\label{sec:synchronisation_numerical_analysis}
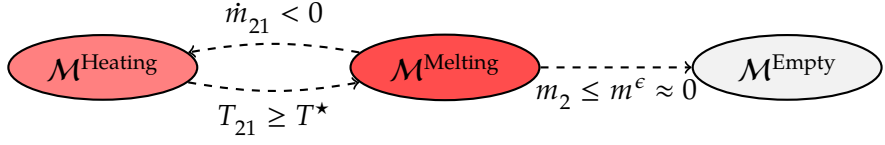
\begin{figure}
	\centering
	\begin{tikzpicture}[node distance=4.5cm]
		\tikzstyle{vertex}=[ellipse, thick, draw=black, text width=1.5cm, align=center]
		\node[vertex, fill=red!50](h) at (0,0) {$\mathcal{M}^{\text{Heating}}$};
 		\node[vertex, fill=red!70, right of=h](m) {$\mathcal{M}^{\text{Melting}}$};
 		\node[vertex, fill=gray!10, right of=m](e) {$\mathcal{M}^{\text{Empty}}$};
		\path (h) edge[->, thick, bend right=10, dashed] node[midway, below] {$\temperature{21} \geq \temperature[\star]{}$} (m);
		\path (m) edge[->, thick, bend right=10, dashed] node[midway, above] {$\mFlowRate{21} < 0$} (h)
				  edge[->, thick, dashed] node[midway, below] {$\mass{2} \leq \mass[\epsilon]{} \approx 0$} (e);
	\end{tikzpicture}
	\caption{Internal states of domain $\domain{2}$ and their transition functions.}
	\label{fig:states_toy_model}
\end{figure}

For instance, let us describe the state of domain $\domain{2}$ into a graph of three states as depicted in \cref{fig:states_toy_model} : \textit{Heating}, \textit{Melting} and \textit{Empty} states. A state transition occurs when a certain algebraic function is activated, e.g. $\mFlowRate{21} < 0$. As explained \cref{sec:coupling_scheme}, each state has its own solver, its own set of equations and its own interface and boundary conditions. State transitions can lead to discontinuities of state or interface variables.
For this case, equations of solver $\mathcal{M}^{\textrm{Heating}}$ are the ones described previously in \cref{sec:discontinuity_numerical_analysis} while solver $\mathcal{M}^{\textrm{Melting}}$ continuous equations are given by
\begin{equation}
\left\{
	\begin{aligned}
		\heatFlux{21} &= \thermalCond{1} \frac{6 \temperature{2} - 4 \temperature{21} - 2 \temperature{b_2}}{\length{2}}\quad\text{in}\;\domain{2}\\
		\D{\mass{2}}{t} &= \massDensity{2}\D{\length{2}}{t} = -\mFlowRate{21}	\quad\text{in}\;\domain{2}\\
		\massDensity{2} \heatCp{2} \length{2} \D{\temperature{2}}{t} &= -\heatFlux{21}\quad\text{in}\;\domain{2},\\
	\end{aligned}
\right.
\end{equation}
leading to a mobile fusion solidification front at interface $\interface{12}$ given by \cref{eq:fusion_front_equilibrium_condition_1,eq:fusion_front_equilibrium_condition_2,eq:fusion_front_equilibrium_condition_3} $\heatFlux{12} = - \heatFlux{21} + \dEnthalpy[\textrm{fus.}]{} \mFlowRate{21}$, $\mFlowRate{12} = -\mFlowRate{21}$ and $\temperature{12} = \temperature{21} = \temperature[\textrm{fus.}]{}$. Again, another set of internal and boundary equations are used for solver $\mathcal{M}^{\textrm{Empty}}$ leading to a fixed interface given by \cref{eq:thermal_equilibrium_condition_1,eq:thermal_equilibrium_condition_2,eq:thermal_equilibrium_condition_3} $\mFlowRate{12} = -\mFlowRate{21} = 0, \temperature{12} = \temperature{21}, \heatFlux{12} = -\heatFlux{21}$. 

Again, at $t=0^-$, both domains are at an equilibrium temperature of $\temperature{1} = \temperature{2} = \temperature{21} = \temperature{b_1} = \temperature{b_2} = 2000$ K. In particular, domain $\domain{2}$ is in \textit{Heating} state. At $t=0^+$, $\temperature{b_1} = 3000$ K and $\temperature{b_2} = 3000$ K to force the \textit{Melting} state transition when $\temperature{21}$ reaches $\temperature[\star]{} = 2100$ K at time $t_{H \rightarrow M}^{\star}$. Once this state has been reached, the second domain is in \textit{Melting} state until only a residual mass $\mass{2} = \mass[\epsilon]{} = 150$ kg is left (corresponding to height $\length{2} = 1.5$ cm) and the \textit{Empty} state is reached at time $t_{M \rightarrow E}^{\star}$. The computation then stops when domain $\domain{1}$ reaches a near stationary state. In addition, all of the computations use the value $\tau_2 = 10^4$ s, $\tau_1 = 8\times10^3$ s and different values for the time step $\Dt$. The chosen values ensure that the \ecs{} is stable, i.e. $r_{12}(\Dt) < 1.0$. \Cref{tab:explicit_implicit_synchro} shows the times of the internal events of domain $\domain{2}$ computed by the \ecs{} and the \ics{} for several coupling time steps. Usually, we ask the coupling time step to be bounded by \[\tau / 10 = 800\ \text{s} \geq \Dt \geq \tau/100 = 80\ \text{s}\quad\text{with}\quad\tau=\min{}(\tau_1,\tau_2)\] for computational efficiency (in red and blue in the table). In comparison to the reference solution (marked with $\dagger$) given by the \ecs{} for a small time step $\Delta t= 1$ s for which the scheme has converged, the \ics{} is able to predict the state transition with high accuracy and adapt its time step to synchronize both domains when the event occurs. However, the synchronization mechanism of the \ecs{} only allows the domains to be informed of an event at the end of the time step. During this window of time, both domains are in a non physical state, i.e.\ the domain $\domain{1}$ is not aware of the disappearance of the domain $\domain{2}$. This leads to numerical errors of order $\mathcal{O}(\Delta t)$ and the numerical creation of either mass and/or energy leading to different stationary states for the \ecs{}. For a time step $\Dt = 100$ s respecting the constraint, the explicit solution leads to a relative error of $18 \%$ in term of mass. Reducing the coupling time step to $10$ s leads to a relative error of $0.3 \%$ but, as seen previously, increases computational time. Besides, the \ics{} always converges toward the reference solution and the right stationary state even for large coupling time step.\\
\begin{table}
\centering
{\renewcommand{\arraystretch}{1.2}
\begin{tabular}{l|c|c|c|c|c|c}
& $\Delta t$ (s) & $\textcolor{blue}{100}$ & $\textcolor{purple}{50}$ & $\textcolor{purple}{25}$ & $\textcolor{purple}{10}$ & $\textcolor{purple}{1}$\\
\hline 
& $r_{12}(\Delta t)$ & \multicolumn{5}{c}{< 1.0}\\
\hline 
\hline
\multirow{2}{*}{\ecs{}} & $\mass[\infty]{1}$ (kg) & $\textcolor{blue}{1492}$ & $\textcolor{purple}{1413}$& $\textcolor{purple}{1347}$ & $\textcolor{purple}{1260}$ & $\textcolor{purple}{1256}\ \dagger$\\
\cline{2-7}
& $\temperature[\infty]{1}$ (K) & $\textcolor{blue}{1710}$ & $\textcolor{purple}{1708}$& $\textcolor{purple}{1705}$ & $\textcolor{purple}{1702}$ & $\textcolor{purple}{1700}\ \dagger$\\
\hline
\multirow{2}{*}{\ics{}} & $\mass[\infty]{1}$ (kg) & $\textcolor{blue}{1255}$ &\multicolumn{4}{c}{$\textcolor{purple}{1255}$}\\
\cline{2-7}
& $\temperature[\infty]{1}$ (K) & $\textcolor{blue}{1700}$ & \multicolumn{4}{c}{$\textcolor{purple}{1700}$}\\
\hline
\hline
\multirow{2}{*}{\ecs{}} & $t_{H \rightarrow M}^{\star}$ (s) & $\textcolor{blue}{1700}$ & $\textcolor{purple}{1600}$& $\textcolor{purple}{1600}$ & $\textcolor{purple}{1590}$ & $\textcolor{purple}{1584}\ \dagger$\\
\cline{2-7}
& $t_{M \rightarrow E}^{\star}$ (s) & $\textcolor{blue}{3200}$ & $\textcolor{purple}{3200}$& $\textcolor{purple}{3150}$ & $\textcolor{purple}{3120}$ & $\textcolor{purple}{3117}\ \dagger$\\
\hline 
\multirow{2}{*}{\ics{}} & $t_{H \rightarrow M}^{\star}$ (s) & $\textcolor{blue}{1583}$ & \multicolumn{4}{c}{$\textcolor{purple}{1583}$}\\
\cline{2-7}
& $t_{M \rightarrow E}^{\star}$ (s) & $\textcolor{blue}{3120}$ & $\textcolor{purple}{3117}$& $\textcolor{purple}{3117}$ & $\textcolor{purple}{3116}$ & $\textcolor{purple}{3116}$\\
\hline
\end{tabular}}
\caption{Times $t_{H \rightarrow M}^{\star}$ and $t_{M \rightarrow E}^{\star}$ of domain $\domain{2}$ internal events and stationary state $\mass[\infty]{1}$ and $\temperature[\infty]{1}$ reached by domain $\domain{1}$. Reference solution marked with $\dagger$, solutions respecting the coupling time step constraint in \textcolor{blue}{blue}, otherwise in \textcolor{purple}{red}.}
\label{tab:explicit_implicit_synchro}
\end{table}

This very simple example of coupling of \lp{} models highlights the main drawback of such lightweight modeling : deleting spatial dependency in the equations force each model to instantly propagate all its data, e.g.\ its discontinuities or its state transitions, thus creating high and important transient or bringing the system in non physical states. Fixes are used to bring back the system in a coherent state which should be avoided at all cost. These phenomenons create numerical errors in the coupling which can be measured in term of numerical energy created at the interface $\interface{12}$. We follow the same methodology as in \cite{piperno_partitioned_2001}. The local variation of energy $\Delta E^{n \rightarrow n+1}$ at interface $\interface{12}$ between times $t^n$ and $t^{n+1}$ is the sum of the energy $\Delta E_1^{n \rightarrow n+1}$ send by domain $\domain{1}$ through the interface and the energy $\Delta E_2^{n \rightarrow n+1}$ send by domain $\domain{2}$ through the interface. For the continuous case in which both domains are strongly coupled, the local variation of energy is null. It is defined by equation
\begin{align*}
\Delta E^{n \rightarrow n+1} &= \Delta E_1^{n \rightarrow n+1}\ -\ (-\Delta E_2^{n \rightarrow n+1})\\ 
&=\int_{t^n}^{t^{n+1}} \heatFlux{12} - \left(\int_{t^n}^{t^{n+1}} \heatFlux{21} - \dEnthalpy[\textrm{fus.}]{} \mFlowRate{21}\right)\\
&= 0.
\end{align*}
The global energy $\Delta E^{0 \rightarrow n}$ through interface $\interface{12}$ is defined by
\begin{equation}
\label{eq:global_energy}
\Delta E^{0 \rightarrow n} = \sum_{k=0}^{n} \Delta E^{k \rightarrow k+1} = 0
\end{equation}
and is also null in the continuous case. For each energy variation $\Delta E$, we define the relative energy
\begin{equation*}
\epsilon(\Delta E) = \frac{\Delta E}{E^{\star}} = \frac{\Delta E}{\mass{2}(t^0)\heatCp{2}(\temperature[\textrm{fus.}]{} - \temperature{2}(t^0)) + \dEnthalpy[\textrm{fus.}]{}\mass{2}(t^0)}
\end{equation*}
by the ratio of the energy $\Delta E$ to the energy $E^{\star}$ needed to heat the initial domain $\domain{2}$ to its fusion temperature ($\mass{2}(t^0)\heatCp{2}(\temperature[\textrm{fus.}]{} - \temperature{2}(t^0))$) and to melt it ($\dEnthalpy[\textrm{fus.}]{}\mass{2}(t^0)$). With the physical values used, we have $E^{\star} \approx 10^8$ J. 

\begin{figure}
    \centering
	\begin{subfigure}[b]{0.5\textwidth}
    	\includegraphics[width=1.0\textwidth]{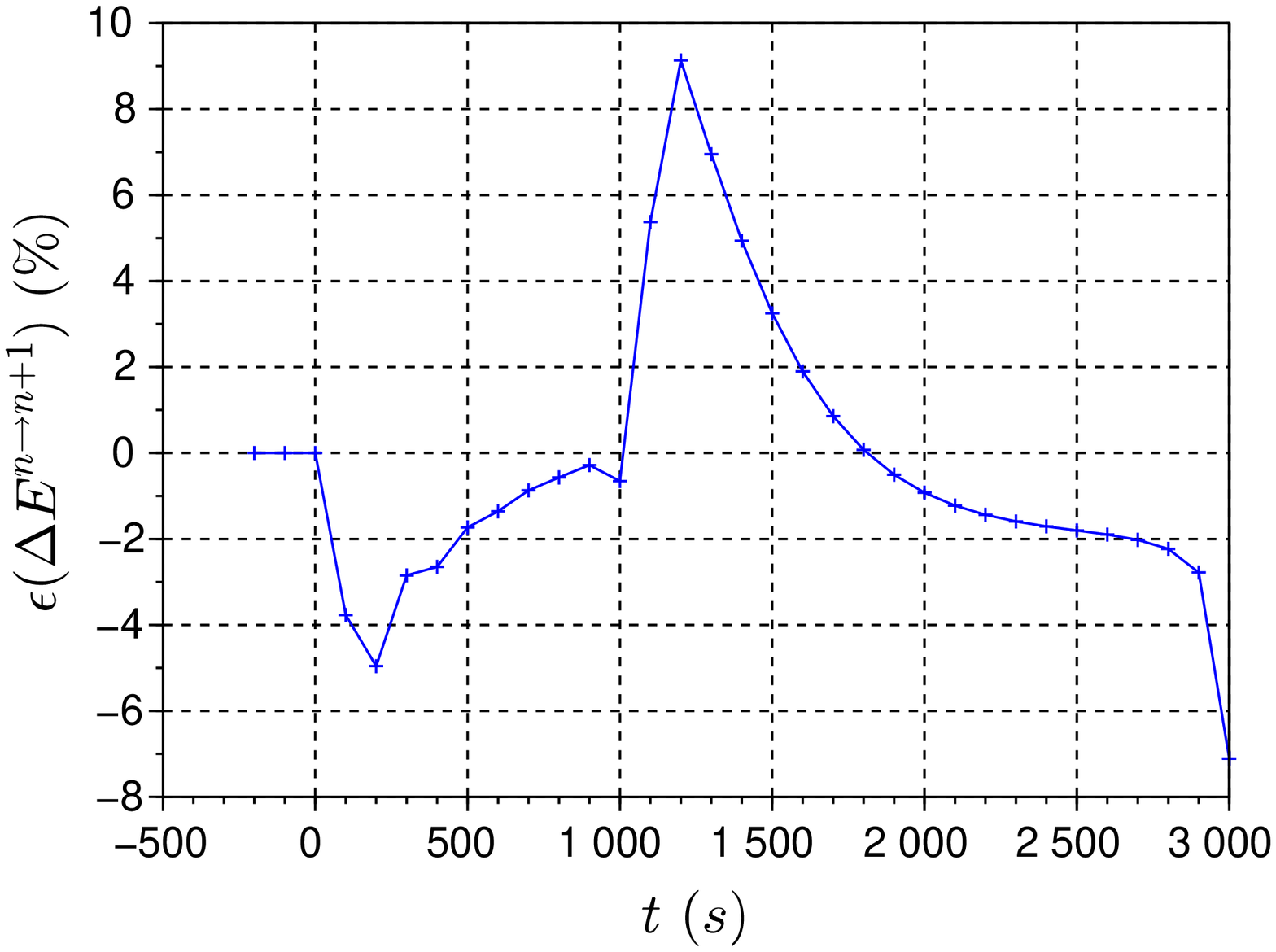}
		\caption{}
		\label{fig:explicit_interface_energy_11}
    \end{subfigure}%
	\begin{subfigure}[b]{0.5\textwidth}
    	\includegraphics[width=1.0\textwidth]{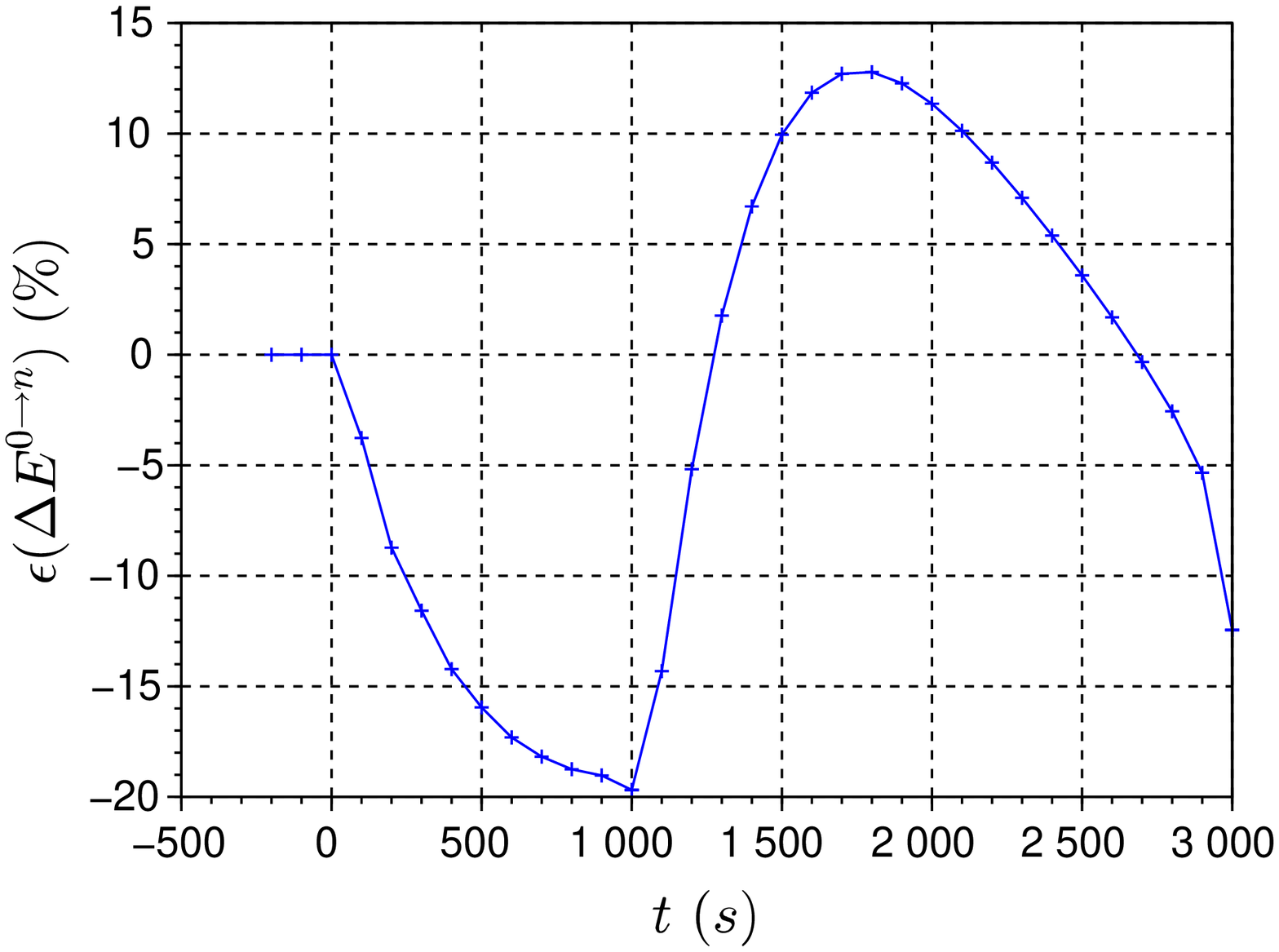}
		\caption{}
		\label{fig:explicit_interface_energy_12}
    \end{subfigure}
	\caption{Relative local energy variation (\textit{left}) and global energy variation (\textit{right}) at interface for the \ecs{}}
	\label{fig:explicit_interface_energy}
\end{figure}

When discretized for the \ecs{}, the local variation of energy becomes 
\begin{equation*}
\Delta E^{n \rightarrow n+1} \approx \heatFlux{12}(t^{n+1})\Delta t\ -\ \left(\heatFlux{21}(t^n)-\dEnthalpy[\textrm{fus.}]{} \mFlowRate{21}(t^n)\right) \Delta t
\end{equation*}
which emphasizes the ``time-lag'' between the two domains. \Cref{fig:explicit_interface_energy_11} shows that the \ecs{} can create locally $2.5 \%$ of $E^{\star}$, i.e. $2.5\,10^6$~J. \Cref{fig:explicit_interface_energy_12} shows first that the global energy is not null and that terms of the sum \cref{eq:global_energy} do not sum to zero. Besides, the explicit scheme can create globally more than $6 \%$ of $E^{\star}$, i.e. $6\,10^6$~J which can have disastrous effects on the whole accuracy.

\begin{figure}
    \centering
	\begin{subfigure}[b]{0.5\textwidth}
    	\includegraphics[width=1.0\textwidth]{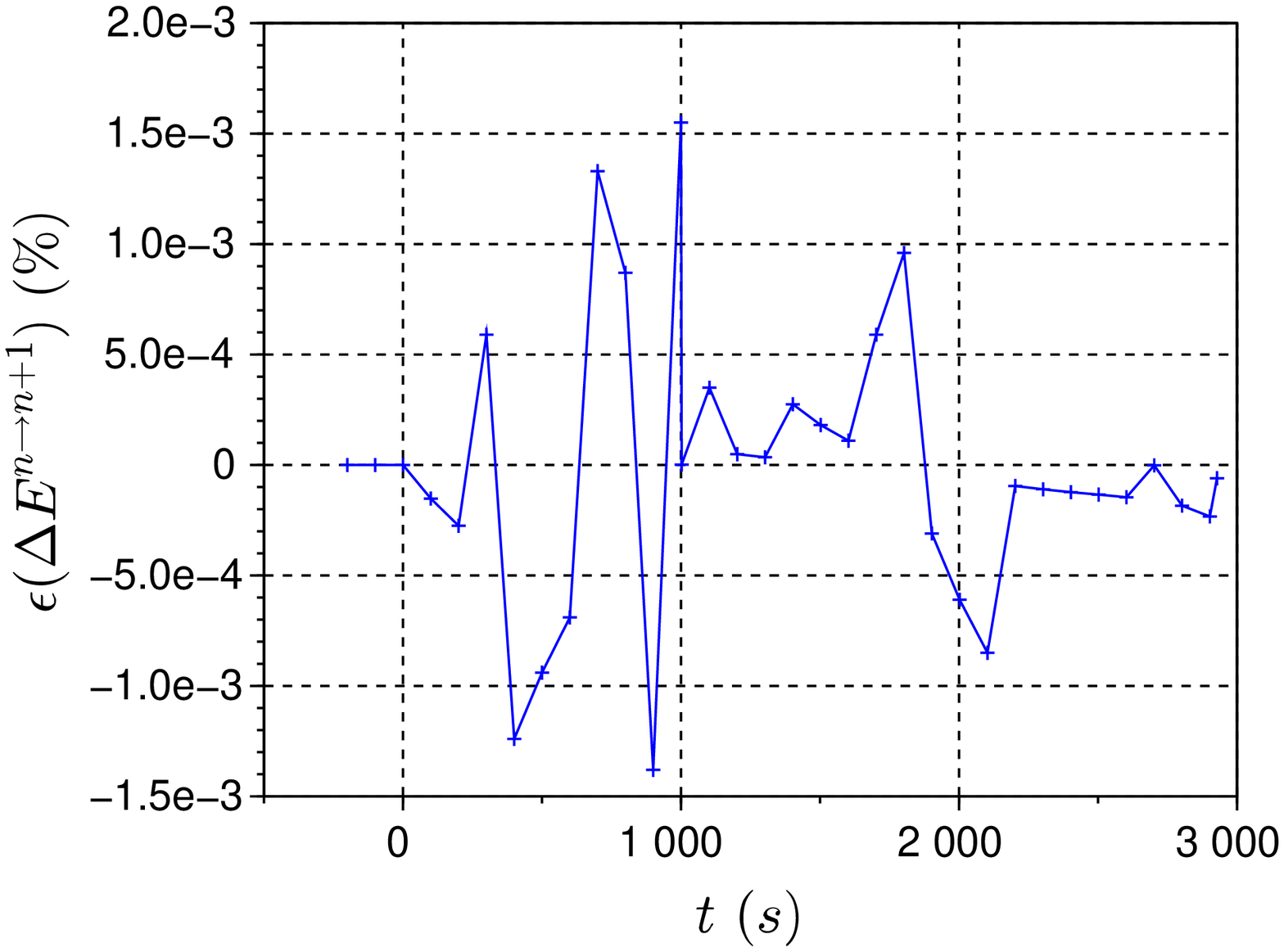}
		\caption{}
		\label{fig:implicit_interface_energy_11}
    \end{subfigure}%
	\begin{subfigure}[b]{0.5\textwidth}
    	\includegraphics[width=1.0\textwidth]{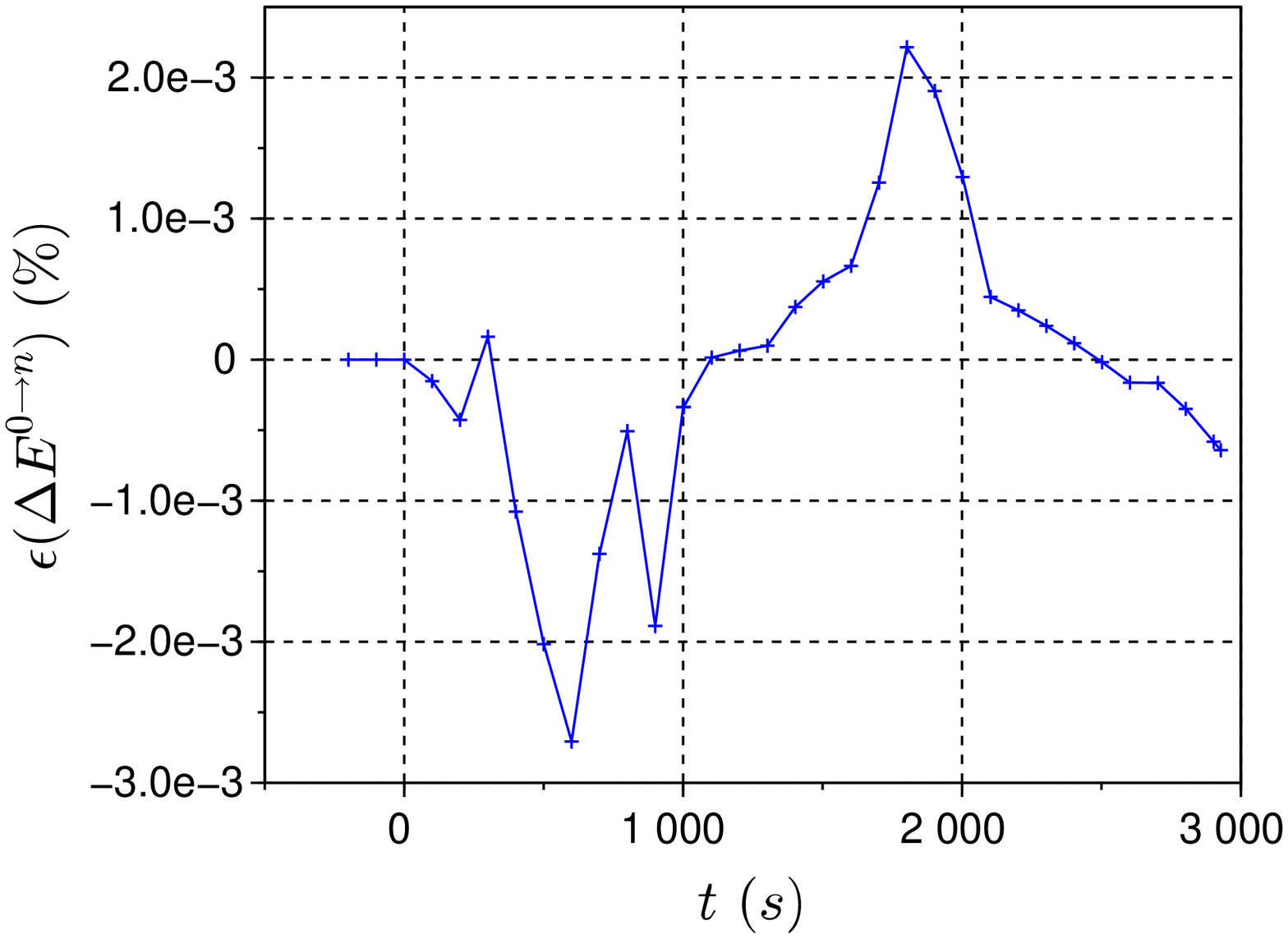}
		\caption{}
		\label{fig:implicit_interface_energy_12}
    \end{subfigure}
	\caption{Relative local energy variation (\textit{left}) and global energy variation (\textit{right}) at interface for the \ics{}}
	\label{fig:implicit_interface_energy}
\end{figure}

For the \ics{}, the local variation of energy becomes 
\begin{equation*}
\Delta E^{n \rightarrow n+1} \approx \heatFlux{12}(t^{n+1,\infty})\Delta t\ -\ \left(\heatFlux{21}(t^{n+1,\infty})-\dEnthalpy[\textrm{fus.}]{} \mFlowRate{21}(t^{n+1,\infty})\right) \Delta t
\end{equation*} 
in which variables with superscript $^\infty$ represent the value reached when the convergence criterion of the iterative scheme is satisfied. At convergence, we can bound the residual at interface in such a way that
\begin{equation*}
\abs{\heatFlux{12}(t^{n+1,\infty})\ -\ \left(\heatFlux{21}(t^{n+1,\infty})-\dEnthalpy[\textrm{fus.}]{} \mFlowRate{21}(t^{n+1,\infty})\right)}\Delta t \leq \epsilon_{\textrm{rel}}\abs{\heatFlux{12}(t^{n+1,\infty})}\Delta t.
\end{equation*} 
This allows us to bound the relative local energy at interface of the \ics{} with
\begin{equation*}
\abs{\epsilon(\Delta E^{n \rightarrow n+1})} = \abs{\frac{\Delta E^{n \rightarrow n+1}}{E^{\star}}} \leq \epsilon_{\textrm{rel}}\ \frac{\abs{\heatFlux{12}(t^{n+1,\infty})}\Delta t}{E^{\star}}
\end{equation*} 
and the relative global energy with
\begin{equation*}
\abs{\epsilon(\Delta E^{0 \rightarrow n})} \leq \sum_{k=0}^{n} \abs{\frac{\Delta E^{k \rightarrow k+1}}{E^{\star}}} \leq n\ \epsilon_{\textrm{rel}}\ \frac{\max\limits_{k}\abs{\heatFlux{12}(t^{k+1,\infty})}\Delta t}{E^{\star}}.
\end{equation*}  
This shows that the imbalance of energy at interface of the \ics{} can be controlled with the relative tolerance $\epsilon_{\textrm{rel}}$. This is confirmed by numerical experiments presented \cref{fig:implicit_interface_energy_11} and \cref{fig:implicit_interface_energy_12} showing respectively the relative local and global energy at interface with a relative tolerance set to $10^{-4}$. The maximum heat flux $\max_{k}\abs{\heatFlux{12}(t^{k+1,\infty})}$ reached at the interface with unit area is equal to $\approx 1.5\times10^5\ \heatFluxUnits{}$ leading to a bound for the relative local energy at interface of $1.5\times10^{-3}$ (expressed in percent in \cref{fig:implicit_interface_energy_11}).

\section{Conclusion and perspectives}
\label{sec:conclusion}
In this paper, we have presented problems of coupled \lp{} models with time management of state change events. In comparison to finer modeling approaches (e.g. mesh based modelings), this approach allows for fast calculations which are needed when doing statistical studies with a large number of calculations (e.g. design of computer experiments, Monte-Carlo methods). This is typically the true industrial context of modeling and simulation of severe accidents in nuclear reactors. However, we have shown that the \lp{} modeling approach often leads to coupled problems with stiff, important and fast transients which are not suitable for being solved by \ecs{}s. Thus, \ics{}s were presented and designed for proper events detection and models synchronization allowing to obtain stable and accurate solutions of coupled problems of lumped parameter models.

We proposed to study theoretically the numerical stability of \ecs{}s and \ics{}s on some of the interface equations used with \lp{} models calculating the heat conduction in coupled domains. From this short study we have learned that Dirichlet-Neumann boundary conditions should be set up carefully in order to ensure stability of the coupling schemes. However with industrial constraints on the coupling time step to ensure fast calculations, \ecs{}s remain very unstable and cannot be reasonable candidates in general cases to give good and precise results. Besides, even the cheapest \ics{}s that only use relaxation can give precise and stable calculations leading to trustworthy results. They turn out ot be very interesting in term of computational times in comparison to \ecs{}s with a highly reduced coupling time step to ensure stability. Furthermore, the designed \ics{}s are able to predict events and discontinuities in the coupled models allowing synchronization between them. This tends to suggest that industrial problems of coupled lumped parameter models could substantially benefit from the use of implicit coupling schemes.\medskip

In future developments, it would be worthwhile to add \emph{smartness} to the coupling scheme. For instance, depending on the strength of a coupling, i.e.\ the value of the residual at interfaces, the coupling scheme sould be able to choose between an \ecs{} or an \ics{} to avoid using potentially costly iterative scheme. Indeed, we have seen that such schemes can be optimized to ensure fast calculations but this is still highly problem dependent. In particular this can be very useful in the context of statistical studies where the experiments are run into an automated process. Finally, if \ics{} are required to ensure proper stability of the computation but are still too costly, time parallelism techniques like \emph{parareal} methods \cite{maday_parareal_2005} could be used. 

\section{Acknowledgments}
This work has been carried out within the framework of the \procor{} platform development funded by CEA, EDF and AREVA.
\bibliographystyle{phdbib}
\bibliography{phdbib}

\begin{thebibliography}{10}

\bibitem{aitken_xxv.bernoullis_1927}
{\textsc A.~C. Aitken}, {\em {XXV}.---{On Bernoulli}'s {Numerical Solution} of
  {Algebraic Equations}}, Proceedings of the Royal Society of Edinburgh, 46
  (1927), pp.~289--305, \url{https://doi.org/10.1017/S0370164600022070},
  \url{https://www.cambridge.org/core/journals/proceedings-of-the-royal-society-of-edinburgh/article/xxvon-bernoullis-numerical-solution-of-algebraic-equations/64D4A7C56F1EFEC696AF68D7870DB451}
  (accessed 2017-10-06).

\bibitem{boccara_modeling_2010}
{\textsc N.~Boccara}, {\em Modeling {Complex Systems}}, Graduate {Texts} in
  {Physics}, Springer New York, New York, NY, 2010,
  \url{https://doi.org/10.1007/978-1-4419-6562-2}.

\bibitem{bonnet_thermohydraulic_1999}
{\textsc J.~M. Bonnet and J.~M. Seiler}, {\em Thermohydraulic phenomena in
  corium pool: the {BALI} experiment}, in Proc. of {ICONE} 7, Tokyo, Japan,
  1999.

\bibitem{causin_added-mass_2005}
{\textsc P.~Causin, J.~F. Gerbeau, and F.~Nobile}, {\em Added-mass effect in
  the design of partitioned algorithms for fluid--structure problems}, Computer
  Methods in Applied Mechanics and Engineering, 194 (2005), pp.~4506--4527,
  \url{https://doi.org/10.1016/j.cma.2004.12.005},
  \url{http://www.sciencedirect.com/science/article/pii/S0045782504005328}
  (accessed 2017-04-25).

\bibitem{degroote_performance_2009}
{\textsc J.~Degroote, K.-J. Bathe, and J.~Vierendeels}, {\em Performance of a
  new partitioned procedure versus a monolithic procedure in fluid--structure
  interaction}, Computers \& Structures, 87 (2009), pp.~793--801,
  \url{https://doi.org/10.1016/j.compstruc.2008.11.013},
  \url{http://linkinghub.elsevier.com/retrieve/pii/S0045794908002605} (accessed
  2016-08-09).

\bibitem{degroote_multi-level_2012}
{\textsc J.~Degroote and J.~Vierendeels}, {\em Multi-level quasi-{Newton}
  coupling algorithms for the partitioned simulation of fluid--structure
  interaction}, Computer Methods in Applied Mechanics and Engineering,
  225–228 (2012), pp.~14--27,
  \url{https://doi.org/10.1016/j.cma.2012.03.010},
  \url{http://www.sciencedirect.com/science/article/pii/S0045782512000862}
  (accessed 2017-03-22).

\bibitem{dolean_introduction_2015}
{\textsc V.~Dolean, P.~Jolivet, and F.~Nataf}, {\em An {Introduction} to
  {Domain Decomposition Methods}}, Other {Titles} in {Applied} {Mathematics},
  Society for Industrial and Applied Mathematics, Nov. 2015,
  \url{https://doi.org/10.1137/1.9781611974065}.

\bibitem{farhat_mixed_1995}
{\textsc C.~Farhat, M.~Lesoinne, and N.~Maman}, {\em Mixed explicit/implicit
  time integration of coupled aeroelastic problems: {Three}-field formulation,
  geometric conservation and distributed solution}, International Journal for
  Numerical Methods in Fluids, 21 (1995), pp.~807--835,
  \url{https://doi.org/10.1002/fld.1650211004/full}.

\bibitem{farhat_unconditionally_1991}
{\textsc C.~Farhat, K.~C. Park, and Y.~Dubois-Pelerin}, {\em An unconditionally
  stable staggered algorithm for transient finite element analysis of coupled
  thermoelastic problems}, Computer methods in applied mechanics and
  engineering, 85 (1991), pp.~349--365,
  \url{http://www.sciencedirect.com/science/article/pii/004578259190102C}
  (accessed 2016-04-01).

\bibitem{farhat_robust_2010}
{\textsc C.~Farhat, A.~Rallu, K.~Wang, and T.~Belytschko}, {\em Robust and
  provably second-order explicit--explicit and implicit--explicit staggered
  time-integrators for highly non-linear compressible fluid--structure
  interaction problems}, International Journal for Numerical Methods in
  Engineering, 84 (2010), pp.~73--107, \url{https://doi.org/10.1002/nme.2883}.

\bibitem{farhat_consistency_1990}
{\textsc C.~Farhat and N.~Sobh}, {\em A consistency analysis of a class of
  concurrent transient implicit/explicit algorithms}, Computer methods in
  applied mechanics and engineering, 84 (1990), pp.~147--162,
  \url{http://www.sciencedirect.com/science/article/pii/0045782590901142}
  (accessed 2017-04-25).

\bibitem{felippa_staggered_1980}
{\textsc C.~A. Felippa and K.~C. Park}, {\em Staggered transient analysis
  procedures for coupled mechanical systems: formulation}, Computer Methods in
  Applied Mechanics and Engineering, 24 (1980), pp.~61--111,
  \url{http://www.sciencedirect.com/science/article/pii/0045782580900407}
  (accessed 2016-08-09).

\bibitem{ganine_nonlinear_2013}
{\textsc V.~Ganine, N.~J. Hills, and B.~L. Lapworth}, {\em Nonlinear
  acceleration of coupled fluid--structure transient thermal problems by
  {Anderson} mixing}, International Journal for Numerical Methods in Fluids, 71
  (2013), pp.~939--959, \url{https://doi.org/10.1002/fld.3689}.

\bibitem{gerbeau_quasi-newton_2003}
{\textsc J.-F. Gerbeau and M.~Vidrascu}, {\em A {Quasi-{Newton} Algorithm
  Based} on a {Reduced Model} for {Fluid-Structure Interaction Problems} in
  {Blood Flows}}, ESAIM: Mathematical Modelling and Numerical Analysis, 37
  (2003), pp.~631--647, \url{https://doi.org/10.1051/m2an:2003049}.

\bibitem{giles_stability_1997}
{\textsc M.~B. Giles}, {\em Stability analysis of numerical interface
  conditions in fluid-structure thermal analysis}, International Journal for
  Numerical Methods in Fluids, 25 (1997), pp.~421--436,
  \url{https://doi.org/10.1002/(SICI)1097-0363(19970830)25:4<421::AID-FLD557>3.0.CO;2-J}.

\bibitem{guillard_significance_2000}
{\textsc H.~Guillard and C.~Farhat}, {\em On the significance of the geometric
  conservation law for flow computations on moving meshes}, Computer Methods in
  Applied Mechanics and Engineering, 190 (2000), pp.~1467--1482,
  \url{https://doi.org/10.1016/S0045-7825(00)00173-0},
  \url{http://www.sciencedirect.com/science/article/pii/S0045782500001730}
  (accessed 2016-08-09).

\bibitem{jacquemain_les_2013}
{\textsc D.~Jacquemain}, {\em Les accidents de fusion du coeur des
  r{\'{e}}acteurs nucl{\'{e}}aires de puissance: {\'{e}}tat des connaissances
  (French)}, Collection sciences et techniques, EDP sciences, Les Ulis, 2013.

\bibitem{joosten_analysis_2009}
{\textsc M.~M. Joosten, W.~G. Dettmer, and D.~Peri{\'{c}}}, {\em Analysis of
  the block {Gauss-Seidel} solution procedure for a strongly coupled model
  problem with reference to fluid-structure interaction}, International Journal
  for Numerical Methods in Engineering, 78 (2009), pp.~757--778,
  \url{https://doi.org/10.1002/nme.2503}.

\bibitem{kassiotis_nonlinear_2011}
{\textsc C.~Kassiotis, A.~Ibrahimbegovic, R.~Niekamp, and H.~G. Matthies}, {\em
  Nonlinear fluid--structure interaction problem. {Part I}: implicit
  partitioned algorithm, nonlinear stability proof and validation examples},
  Computational Mechanics, 47 (2011), pp.~305--323,
  \url{https://doi.org/10.1007/s00466-010-0545-6}.

\bibitem{kuttler_fixed-point_2008}
{\textsc U.~K{\"{u}}ttler and W.~A. Wall}, {\em Fixed-point fluid--structure
  interaction solvers with dynamic relaxation}, Computational Mechanics, 43
  (2008), pp.~61--72, \url{https://doi.org/10.1007/s00466-008-0255-5}.

\bibitem{le_tellier_phenomenological_2015}
{\textsc R.~Le~Tellier, L.~Saas, and F.~Payot}, {\em Phenomenological analyses
  of corium propagation in {LWRs}: the {PROCOR} software platform}, in Proc. of
  the 7\textsuperscript{th} {European} {Review} {Meeting} on {Severe}
  {Accident} {Research} {ERMSAR}-2015, Marseille, France, 2015.

\bibitem{le_tellier_treatment_2017}
{\textsc R.~Le~Tellier, E.~Skrzypek, and L.~Saas}, {\em On the treatment of
  plane fusion front in lumped parameter thermal models with convection},
  Applied Thermal Engineering, 120 (2017), pp.~314--326,
  \url{https://doi.org/10.1016/j.applthermaleng.2017.03.108},
  \url{http://www.sciencedirect.com/science/article/pii/S1359431116327119}
  (accessed 2017-04-10).

\bibitem{maday_parareal_2005}
{\textsc Y.~Maday and G.~Turinici}, {\em The Parareal in Time Iterative Solver:
  a Further Direction to Parallel Implementation}, Springer Berlin Heidelberg,
  Berlin, Heidelberg, 2005, pp.~441--448,
  \url{https://doi.org/10.1007/3-540-26825-1_45}.

\bibitem{mao_efficient_2002}
{\textsc G.~Mao and L.~R. Petzold}, {\em Efficient integration over
  discontinuities for differential-algebraic systems}, Computers \& Mathematics
  with Applications, 43 (2002), pp.~65--79,
  \url{https://doi.org/10.1016/S0898-1221(01)00272-3},
  \url{http://www.sciencedirect.com/science/article/pii/S0898122101002723}
  (accessed 2017-06-21).

\bibitem{mehl_parallel_2016}
{\textsc M.~Mehl, B.~Uekermann, H.~Bijl, D.~Blom, B.~Gatzhammer, and A.~van
  Zuijlen}, {\em Parallel coupling numerics for partitioned fluid--structure
  interaction simulations}, Computers \& Mathematics with Applications, 71
  (2016), pp.~869--891, \url{https://doi.org/10.1016/j.camwa.2015.12.025},
  \url{http://linkinghub.elsevier.com/retrieve/pii/S0898122115005933} (accessed
  2016-08-09).

\bibitem{michler_interface_2005}
{\textsc C.~Michler, E.~H. van Brummelen, and R.~de~Borst}, {\em An interface
  {Newton-Krylov} solver for fluid-structure interaction}, International
  Journal for Numerical Methods in Fluids, 47 (2005), pp.~1189--1195,
  \url{https://doi.org/10.1002/fld.850}.

\bibitem{minami_performance_2010}
{\textsc S.~Minami and S.~Yoshimura}, {\em Performance evaluation of nonlinear
  algorithms with line-search for partitioned coupling techniques for
  fluid-structure interactions}, International Journal for Numerical Methods in
  Fluids, 64 (2010), pp.~1129--1147, \url{https://doi.org/10.1002/fld.2274}.

\bibitem{petzold_differential/algebraic_1982}
{\textsc L.~Petzold}, {\em Differential/{Algebraic Equations} are not
  {ODE\textquoteright}s}, SIAM Journal on Scientific and Statistical Computing,
  3 (1982), pp.~367--384, \url{https://doi.org/10.1137/0903023}.

\bibitem{piperno_partitioned_2001}
{\textsc S.~Piperno and C.~Farhat}, {\em Partitioned procedures for the
  transient solution of coupled aeroelastic problems--{Part II}: energy
  transfer analysis and three-dimensional applications}, Computer methods in
  applied mechanics and engineering, 190 (2001), pp.~3147--3170,
  \url{http://www.sciencedirect.com/science/article/pii/S0045782500003868}
  (accessed 2016-08-09).

\bibitem{piperno_partitioned_1995}
{\textsc S.~Piperno, C.~Farhat, and B.~Larrouturou}, {\em Partitioned
  procedures for the transient solution of coupled aroelastic problems {Part
  I}: {Model} problem, theory and two-dimensional application}, Computer
  Methods in Applied Mechanics and Engineering, 124 (1995), pp.~79 -- 112,
  \url{https://doi.org/10.1016/0045-7825(95)92707-9},
  \url{http://www.sciencedirect.com/science/article/pii/0045782595927079}.

\bibitem{ramiere_iterative_2015}
{\textsc I.~Rami{\`{e}}re and T.~Helfer}, {\em Iterative residual-based vector
  methods to accelerate fixed point iterations}, Computers \& Mathematics with
  Applications, 70 (2015), pp.~2210--2226,
  \url{https://doi.org/10.1016/j.camwa.2015.08.025},
  \url{http://linkinghub.elsevier.com/retrieve/pii/S0898122115004046} (accessed
  2017-01-16).

\bibitem{sehgal_nuclear_2012}
{\textsc B.~R. Sehgal}, {\em Nuclear safety in {Light Water Reactors}: {Severe
  Accident Phenomenology}}, Elsevier/Academic Press, Amsterdam ; Boston, 1st
  ed~ed., 2012.

\bibitem{vierendeels_implicit_2007}
{\textsc J.~Vierendeels, L.~Lanoye, J.~Degroote, and P.~Verdonck}, {\em
  Implicit coupling of partitioned fluid--structure interaction problems with
  reduced order models}, Computers \& Structures, 85 (2007), pp.~970--976,
  \url{https://doi.org/10.1016/j.compstruc.2006.11.006},
  \url{http://www.sciencedirect.com/science/article/pii/S0045794906003865}
  (accessed 2017-05-05).

\bibitem{zhang_natural_2015}
{\textsc L.~Zhang, Y.~Zhou, Y.~Zhang, W.~Tian, S.~Qiu, and G.~Su}, {\em Natural
  convection heat transfer in corium pools: {A} review work of experimental
  studies}, Progress in Nuclear Energy, 79 (2015), pp.~167--181,
  \url{https://doi.org/10.1016/j.pnucene.2014.11.021},
  \url{http://www.sciencedirect.com/science/article/pii/S014919701400328X}
  (accessed 2016-10-06).

\end{thebibliography}
\end{document}